\documentclass[reprint, secnumarabic, amssymb, nobibnotes, aps, pra,superscriptaddress]{revtex4-1}
\usepackage[utf8]{inputenc}
\usepackage[left=1in,right=1in,top=2cm,bottom=2cm]{geometry}
\usepackage{graphicx,lipsum,amsmath}

\usepackage{xcolor}

\begin{document}

\title{Phases of interacting bosons in a hybrid Harper-Hofstadter system with a synthetic dimension of harmonic trap states}
\author{David G. Reid}
\affiliation{School of Physics and Astronomy, University of Birmingham, Edgbaston, Birmingham B15 2TT, United Kingdom}
\author{Holly A. J. Middleton-Spencer}
\affiliation{School of Physics and Astronomy, University of Birmingham, Edgbaston, Birmingham B15 2TT, United Kingdom}
\author{Grazia Salerno}
\affiliation{Department of Applied Physics, School of Science, Aalto University, Espoo, Finland}
\affiliation{Dipartimento di Fisica ``E. Fermi’’, Università di Pisa, Largo Bruno Pontecorvo 3, Pisa, 56127 Italy}
\author{Nathan Goldman}
\affiliation{International Solvay Institutes, 1050 Brussels, Belgium}
\affiliation{Center for Nonlinear Phenomena and Complex Systems, Université Libre de Bruxelles (U.L.B.), B-1050 Brussels, Belgium}
\affiliation{Laboratoire Kastler Brossel, Coll\`ege de France, CNRS, ENS-Universit\'e PSL, Sorbonne Universit\'e, 11 Place Marcelin Berthelot, 75005 Paris, France}
\author{Hannah M. Price}
\affiliation{School of Physics and Astronomy, University of Birmingham, Edgbaston, Birmingham B15 2TT, United Kingdom}

\begin{abstract}
{Synthetic dimensions are a powerful tool for engineering desired quantum systems, based on coupling together sets of states and reinterpreting these as lattice sites along an artificial dimension. Recently, a synthetic dimension of harmonic trap states has been successfully implemented in an ultracold atom experiment, opening the way for future realizations in this platform of topological lattice models, such as hybrid Harper-Hofstadter (HH) systems, which have one real and one synthetic dimension. However, unlike conventional systems, inter-particle interactions along a synthetic dimension of harmonic trap states are inhomogeneous, long-ranged and non-state-preserving. Therefore, this setup provides a natural platform for the exploration of the interplay between long range interactions (including correlated pair tunneling) and magnetic effects. In this paper, we set out to numerically study the effect of such interactions on both a hybrid two-legged HH ladder and a 2D HH model. In the former, we find variants of vortex and Meissner phases familiar from conventional models, while in the latter, we observe the emergence, in small finite systems, of unusual ground states, including a ``Meissner stripe" state, which combines counter-propagating Meissner-like currents with strong density variations. This opens up interesting questions, including about the nature of strongly-correlated states that would emerge in such a platform.}
\end{abstract}
\maketitle

  \section{Introduction}

  Over the last decade, synthetic dimensions have become established as a powerful technique for the quantum simulation of lattice physics~\cite{Boada2012, Celi2014, Mancini2015, Stuhl2015, Livi2016, Kolkowitz2017, Roell2023, Chalopin2020, lu2024probing, lienhard2020realization,Kanungo2021, ozawa2017synthetic, Price2017, Gadway2016, Viebahn2019, PRXQuantum.5.010310, bouhiron2024realization, cai2019experimental,Sundar2018, kang2020creutz,ozawa2019topological, lustig2019photonic,dutt2020single,balvcytis2021synthetic,chen2021real,lustig2021topological,oliver2023bloch,oliver2023artificial,dinh2024reconfigurable,leefmans2022topological,balvcytis2022synthetic,ehrhardt2023perspective,PhysRevLett.132.130601,lauria2022}. In this approach, a set of internal states or degrees of freedom are coupled together so that they can be re-interpreted as lattice sites for a single particle hopping along an artificial spatial dimension. This has led to the experimental exploration of many interesting phenomena, including quantum Hall physics in both two~\cite{Celi2014, Mancini2015, Stuhl2015,lustig2019photonic,Chalopin2020,Gadway2016,PhysRevLett.132.130601} and even four dimensions~\cite{Viebahn2019, Boada2012,price2015, bouhiron2024realization}, non-Hermitian topological bands~\cite{Wang2021} and Anderson physics~\cite{meier2018observation,li2022aharonov,zeng2024transition}.

The current frontier for experiments is now to include (strong) inter-particle interactions to use synthetic dimensions for exploring many-body physics, such as strongly-correlated topological states~\cite{ozawa2019topological,fabre2024atomic}. However, the inter-particle interactions along the synthetic dimension are typically very different to those familiar from real spatial dimensions as particles that are far apart in the synthetic dimension are often physically close in real space and so still interact strongly. Hybrid systems, with both synthetic and real spatial dimensions,  then typically also have very anisotropic interactions, unlike in conventional condensed matter models. Furthermore, interactions do not necessarily conserve the position along the synthetic dimension if there are state-changing collisions between particles~\cite{chang2005coherent, Price2017}. 

The nature of the synthetic-dimension interactions, and hence the resulting many-body physics, therefore changes dramatically depending on what states are used to simulate the synthetic lattice. For example, using the hyperfine states of alkaline earth atoms can lead to effectively $SU(N)$-invariant interactions along the synthetic dimension, where $N$ refers to the number of hyperfine states involved~\cite{Celi2014,Mancini2015}. These interactions are thus independent of the position along the synthetic dimension, and hence are of ``infinite range". Consequences of these interactions has been extensively studied theoretically~\cite{barbarino2015magnetic, yan2015topological,cornfeld2015chiral,barbarino2016synthetic,zeng2015charge,taddia2017topological,junemann2017exploring,ghosh2015baryon,mamaev2022resonant,barbiero2023frustrated,buser2020interacting} and recently experimentally~\cite{zhou2023observation} in hybrid ladder geometries spanned by a few synthetic sites; however, it still remains to be seen whether these interactions will lead to interesting fractional quantum Hall-like states in larger lattices~\cite{calvanese2017laughlin,saito2017devil} or if other techniques must be employed to reduce the effective range of interactions~\cite{barbiero2020bose, Chalopin2020}. Other recent synthetic-dimension experiments in cold atoms have explored interacting physics using momentum states, which give rise to effectively local attractive interactions in the synthetic dimension~\cite{an2018correlated, An2021,wang2022observation}, and with Rydberg states, where the strong dipolar interactions between atoms leads to correlated tunneling effects along the synthetic dimension~\cite{chen2024interaction, chen2024strongly}, which may support exotic quantum string and membrane many-body phases in larger systems~\cite{Sundar2018,Sundar2019}. 

In this work, we theoretically explore the effects of inter-particle interactions along a synthetic dimension of atomic trap states that are coupled together by periodically shaking the harmonic trapping potential~\cite{Price2017, Salerno2019, reid}. Compared to other schemes, this approach has the advantage that it can be straightforwardly implemented in many platforms, as it requires few experimental ingredients, and it naturally leads to a very long and controllable synthetic dimension, consisting of many tens of sites, as recently shown experimentally~\cite{oliver2023bloch}. Interestingly, it is as yet unknown what the effects of inter-particle interactions will be in this scheme as even contact interactions in real-space translate into exotic interaction processes in synthetic space, including state-changing collisions as well as interactions that depend on both the absolute and relative positions of two particles along the trap-state dimension~\cite{Price2017}. The interactions are also inherently long-ranged along the harmonic synthetic dimension meaning that this set-up can provide a simple and natural setting in which to explore the interplay between long-range interactions and magnetic-flux effects.  

To begin addressing this problem, we have studied how  inter-particle interactions change the mean-field bosonic condensate phases of both the two-legged Harper-Hofstadter ladder (HH) and the full two-dimensional HH model, describing particles hopping on a square ladder or lattice, respectively, in the presence of a perpendicular magnetic field~\cite{Atala_2014, harper1955single,hofstadter1976energy, natu2015bosons}. These effects can be simply implemented in cold atom systems via the use of artificial gauge fields \cite{Dalibard_2011, Galitski_2019}.
We have numerically explored what happens if one spatial dimension of each HH system is replaced with a synthetic lattice of atomic trap states, as may be implemented in future experiments~\cite{Price2017}. Firstly, for the two-legged HH ladder, we identify variants of the usual Meissner and vortex-lattice phases~\cite{Atala_2014}, as will be further introduced below. Secondly, we find, for the 2D HH model, that the synthetic-dimension interactions dramatically change the ground state, leading to the emergence of features unlike those seen with conventional interactions~\cite{PhysRevA.83.013612}. This includes the appearance of a new type of state, particularly for small synthetic-dimension lengths, which we designate as the ``Meissner stripe" state as it combines Meissner-like currents along the real dimension with strong density modulations along the synthetic dimension. In the future, it will be interesting to push this work further towards making a realistic experimental proposal for how to prepare these states, and to go beyond mean-field interactions to stronger interaction strengths where the usual HH model hosts lattice versions of fractional quantum Hall states, as also recently observed experimentally in a small real-space lattice with ultracold atoms~\cite{leonard2023realization}.     
  
\subsection*{Outline}
The outline of the paper is as follows: in Section~\ref{Sec:scheme} we briefly introduce Harper-Hofstadter physics, before reviewing how this can be realized using a synthetic dimension of atomic trap states. We then review the properties of inter-particle contact interactions in the synthetic-dimension picture, and summarize the main methods and observables studied. In Section~\ref{sec:ladder}, we study the ground states associated with a two-legged HH ladder, comparing to known results from the literature. In Section~\ref{sec:vortex_ground_states}, we extend these systems to investigate the ground states of the 2D HH model for different magnetic flux strengths and present the exotic states these interactions induce, including a Messier-stripe phase which we will detail in this section. Additionally, within this Section, we will compare these new phases with results from standard mean-field contact interactions, previously studied in Ref.~\cite{PhysRevA.83.013612} and the results from the ladder investigation (Section~\ref{sec:ladder}). Finally, in Section~\ref{sec:conclusions}, we discuss our results and draw conclusions. 

\section{Scheme}
\label{Sec:scheme}

\subsection{Harper-Hofstadter Physics}

In this Section we begin by briefly introducing the Harper-Hofstadter model, which describes a charged particle hopping on a square lattice in the presence of a perpendicular magnetic field as~\cite{harper1955single, hofstadter1976energy} 
\begin{eqnarray}
\mathcal{H}_{\text{HH}} = -\sum_{n, \lambda} &&\left[ t e^{-i\varphi n } \psi^{\dagger}_{n, \lambda-1} \psi_{n, \lambda} \right. \nonumber \\
&& \left. + J_x\psi^{\dagger}_{n-1, \lambda}\psi_{n, \lambda} + \text{h.c.} \right], 
\label{Eq:HH}
\end{eqnarray}
where $\psi_{n, \lambda}$ ($\psi^\dagger_{n, \lambda} $) annihilates (creates) a particle at site $(n, \lambda)$, and where $J_x$ and $t$ are the hopping amplitudes along the two directions. Here, we have chosen the Landau gauge where the hopping along the $\lambda$ direction is modified by a complex phase $\varphi n$, such that there is a uniform magnetic flux $\Phi\!=\!\varphi$ piercing each plaquette. Note that here we have set $q\!=\!\hbar \!=\!1$. In this paper, we shall also often define $\varphi\equiv2 \pi \alpha$, such that $\alpha$ corresponds to the number of magnetic flux quanta per plaquette.

The HH model has been widely studied as it is a canonical model for 2D quantum Hall physics with a single-particle energy band-structure characterized by non-zero topological Chern numbers and one-way chiral edge states. Thanks to experimental advances, such as the development of artificial gauge fields, this model has been experimentally investigated in many different physical set-ups, including with cold atoms in optical lattices~\cite{aidelsburger, Cooper2019} and with a synthetic dimension of hyperfine states~\cite{Chalopin2020}.    

The HH ladder is then the quasi-1D limit of the above HH model [Eq.~(\ref{Eq:HH})] in which there are only a few sites along either $n$ or $\lambda$, while the other direction remains extended. This is therefore a minimal model in which to explore the hallmarks of magnetic fields and topology in reduced dimensions. In particular, cold-atom experiments have previously used both real~\cite{Atala_2014} and synthetic dimensions~\cite{Mancini2015, Stuhl2015} to realize this model and study phenomena such as magnetic skipping orbits and chiral currents. Recent experiments have also pushed these systems further to the strongly-interacting regime, where interesting many-body states have been explored~\cite{leonard2023realization, impertro2025strongly}.

 In this paper, we will be focusing on the case of bosons in the weakly-interacting (mean-field) limit; in this regime, the HH ladder usually exhibits a rich phase diagram, including so-called vortex-lattice, Meissner and biased-ladder phases~\cite{wei2014theory, Atala_2014,piraud2015vortex, kelecs2015mott,qiao2021quantum,uchino2015population,natu2015bosons,kolovsky2017bogoliubov}, as will be reviewed in more detail below [c.f. Sec.~\ref{sec:ladder}]. In a similar regime, the superfluid phase of the HH 2D square lattice exhibits density modulations and chiral currents~\cite{PhysRevA.83.013612} [c.f. Sec.~\ref{sec:vortex_ground_states}]. These condensate configurations are similar to structures found in superconducting lattices in magnetic fields~\cite{alexander1983superconductivity}, and, in the low-field limit, resemble Abrikosov vortex lattices. Our aim will be to investigate what happens to these phases if we include the effects of (exotic) interactions along the synthetic dimension of atomic trap states [c.f. Sec.~\ref{sec:inter}].

\subsection{Engineering HH Physics with a Synthetic Dimension of Trap States}

\begin{figure}
\centering
\includegraphics[width=\columnwidth]{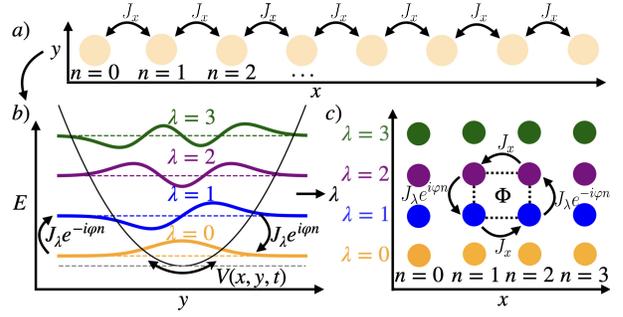}
\caption{\label{fig:schematic}  A schematic for how to engineer HH physics. (a) Along the $x$ direction, there is a single-band tight-binding lattice, with hopping amplitude $J_x$ and with sites indexed by $n$. (b) Along the $y$ direction, there is a strong harmonic trap with states indexed by $\lambda$. Applying the periodic potential $V(x, y, t)$ [Eq.~(\ref{Eq:realistic_potential})] leads to an effective Floquet Hamiltonian [Eq.~(\ref{Eq:F})]  in which neighboring trap states are coupled with an amplitude $J_{\lambda} $ and a phase that depends on the position along the $x$ direction. (c) When viewed in the hybrid synthetic-real $\lambda\!-\!x$ plane, this is a Harper-Hofstadter model for a charged particle hopping in the presence of a magnetic flux $\Phi$, controlled by the hopping phase $\varphi$.    
}
\end{figure}

We shall now briefly review how HH-like physics may be engineered using a synthetic dimension of atomic trap states~\cite{price2015, Salerno2019}. Note that for simplicity, we shall afterwards focus in the main text on the ideal HH equation described above [Eq.~(\ref{Eq:HH})], so readers may proceed directly to Sec.~\ref{sec:inter} if desired. However, this section does explain the physical meaning and approximations behind this synthetic-dimension approach, which in turn helps to understand the origin of the interaction effects to be studied. 

To engineer a synthetic dimension of trap states, we consider starting from an atomic gas in an elongated ``cigar-shaped" harmonic trap with trapping frequencies along the $x$, $z$ and $y$ direction chosen as, respectively, $\omega_x \ll  \omega_z, \omega_y $~\cite{Price2017,oliver2023bloch,Salerno2019,reid}. Henceforth, we will assume that the motion along $z$ can be neglected due to the strong trapping potential in this direction, allowing us to restrict our discussion to the $x\!-\!y$ plane. Along the $x$ direction, we then  impose a deep optical lattice potential such that the corresponding motion is described in the single-band tight-binding approximation. The single-particle part of the Hamiltonian is then given by~\cite{Salerno2019}
\begin{equation}
    \hat{H}_0 = \frac{\hat{p}_y^2}{2M} + \frac{M\omega_y^2}{2} y^2 - J_x \sum_n (|y,n\rangle \langle y,n-1| + \text{h.c.}).
    \label{eq:hamiltonian}
\end{equation}
where $M$ is the particle mass, $J_x$ is the hopping amplitude and where $n\!=\!0, 1, 2, ..., (N_{n}-1)$ indexes the optical lattice sites with $N_n$ being the total number of sites, as sketched in Fig.~\ref{fig:schematic}a.

To realise the synthetic dimension, we add a suitable time-periodic driving potential such as 
\begin{equation}
V(x, y, t) = -V_0 \left( 2 \Theta \left(y \right) - 1\right) \cos ( \omega_D t + \varphi n),
\label{Eq:realistic_potential}
\end{equation}
where $\Theta(...)$ is the Heaviside function and where $V_0$ and $\omega_D$ are the driving strength and frequency respectively. We choose the driving potential to have this spatial dependence along $y$ as it  is simple to implement experimentally, for example, via a digital micromirror device, and it results in effective hopping amplitudes close to the theoretical idealized case. Hereafter, we assume that the driving is on resonance with the trapping frequency $\omega_D\! = \!\omega_y\! \equiv \! \omega $; introducing a detuning leads to an effective force along the synthetic dimension, as recently experimentally explored in Ref.~\cite{oliver2023bloch}. Importantly, we also note that this modulation depends explicitly on the position in the real-space optical lattice through its phase $\phi(x)=\varphi n$. Such a modulation potential could be created experimentally, for example, by using a digital micromirror device, similar to in the recent experiment~\cite{oliver2023bloch}.

An effective stroboscopic Floquet model can then be obtained from the time-dependent Hamiltonian, $\hat{H}=\hat{H}_0+V$, under a rotating-wave approximation when $\omega$ is much larger than all other frequency scales,  as~\cite{rudner2020floquetengineershandbook, PhysRevX.4.031027,Price2017} 
\begin{eqnarray}
\mathcal{H}_{\text{F}} = \sum_{n, \lambda} &&\left[\epsilon_\lambda\psi_{n,\lambda}^\dagger \psi_{n,\lambda}+ J_{\lambda} e^{-i\varphi n } \psi^{\dagger}_{n, \lambda-1} \psi_{n, \lambda} \right. \nonumber \\
&& \left. -J_x\psi^{\dagger}_{n-1, \lambda}\psi_{n, \lambda} + \text{h.c.} \right],
\label{Eq:F}
\end{eqnarray}
 where $\lambda \!=\! 0, 1, 2,..., (N_\lambda-1)$ indexes the atomic trap states [c.f.~Fig.~\ref{fig:schematic}b], with $N_\lambda$ being the total number of trap states considered, $\epsilon_{\lambda}$ acts as an onsite potential and  $J_{\lambda}$ being the effective hopping amplitude along the synthetic dimension.  Within this scheme, the number of $\lambda$ states described by Eq.~(\ref{Eq:F}) is typically limited by anharmonicities in the trapping potential, allowing for the creation of long synthetic dimensions with, for example, tens of sites as in the recent experiment~\cite{oliver2023bloch}. However, it is also possible to add additional confining potentials to the ultracold gas so as to sharply cut off all states above a desired threshold in $\lambda$, by tuning them away from resonance, meaning that the total number of states, $N_\lambda$, can be easily reduced~\cite{reid}.    
 
 The effective Floquet Hamiltonian [Eq.~(\ref{Eq:F})] is clearly a variant of the 2D HH model [Eq.~(\ref{Eq:HH})] with an onsite potential and non-uniform hopping along the $\lambda$ direction. In fact, we have chosen the form of the modulation [Eq.~(\ref{Eq:realistic_potential})] such that $J_{\lambda}$ oscillates around a constant value, $-t$, with the magnitude of oscillations rapidly decaying with increasing $\lambda$ and $\epsilon_{\lambda}$ is essentially constant for allowed values of $\lambda$ (cf. Appendix A). For this reason, we hereafter approximate the hoppings as being uniform, $J_{\lambda}\!=\! -t$, and treat the onsite potential as a constant offset allowing it to be scaled out, i.e. $\epsilon_{\lambda}=0$. We instead work directly with Eq.~(\ref{Eq:HH}).
 Note that we have checked that this assumption of uniform hoppings does not significantly affect our results [see Appendix D]. If alternative modulation potentials are used then $J_\lambda$ can exhibit a stronger dependence on $\lambda$; as an example, we briefly discuss the modulation used in the experiment of Ref.~\cite{oliver2023bloch} in Appendix E, where we again observe that this does not significantly change our results. 

 Finally, to explore the physics of a HH ladder, we need to restrict the number of sites in either the $x$ or $\lambda$ direction, as mentioned above. Physically, we can realize such a restriction in the $x$ direction by replacing the deep optical lattice with, for example, a double-well potential or superlattice. Instead to restrict the number of $\lambda$ sites, we can modify the harmonic trap potential along $y$ in Eq.~(\ref{eq:hamiltonian}) so that the applied modulation only effectively couples a few states; this can be achieved, for example, by adding a square-well potential along $y$, in order to tune all trap states above a certain $\lambda$ away from resonance~\cite{reid}. Note that while the single-particle physics is the same regardless of whether the long axis of the ladder is aligned along the $x$ or $\lambda$ directions, the many-body physics will be very different due to the unusual form of the inter-particle interactions with respect to the synthetic dimension. Hereafter we specify the orientation of our two-leg HH ladder such that the extended ``leg" direction is along the synthetic dimension, while the short ``rung" direction is along the real $x$ direction with the double well configuration; this is because we are interested in finding novel effects due to the synthetic-dimension interactions, and so want to maximize the number of different $\lambda$ states involved.

\subsection{Interactions along a synthetic dimension of harmonic trap states} \label{sec:inter}

We will now review how contact interactions between particles in real space become long-ranged along the synthetic dimension of atomic trap states~\cite{Price2017}. We begin with the $2$D real-space contact interaction term
\begin{equation}
\mathcal{H}_{\text{int}}  = \frac{g}{2}\sum_n\int   \psi_{n}^{\dagger}(y)\psi_{n}^{\dagger}(y)  \psi_{n}(y)\psi_{n}(y) dy,
\end{equation} 
where $g$ is the interaction strength and $\psi_{n}(y)$ ($\psi_{n}^{\dagger}(y)$) annihilates (creates) a particle at site $n$ and position $y$. This can be mapped into synthetic space by projecting the wave-function onto the harmonic trap eigenstates by
\begin{eqnarray}
\psi_n (y) = \frac{1}{\sqrt{l_H}}\sum_{\lambda} h_{\lambda}(y)a_{n, \lambda},
\label{Eq:intstep1}
\end{eqnarray} 
where $a_{n,\lambda}$ ($a^\dagger_{n, \lambda}$) annihilates (creates) a particle at site $n$ and state $\lambda$ (in the non-rotating frame), $l_H =1/\sqrt{M\omega_y}$ is the harmonic oscillator length and $h_{\lambda}(y)$ is the normalized Hermite polynomial
\begin{figure*}
\centering
\includegraphics[width=\textwidth]{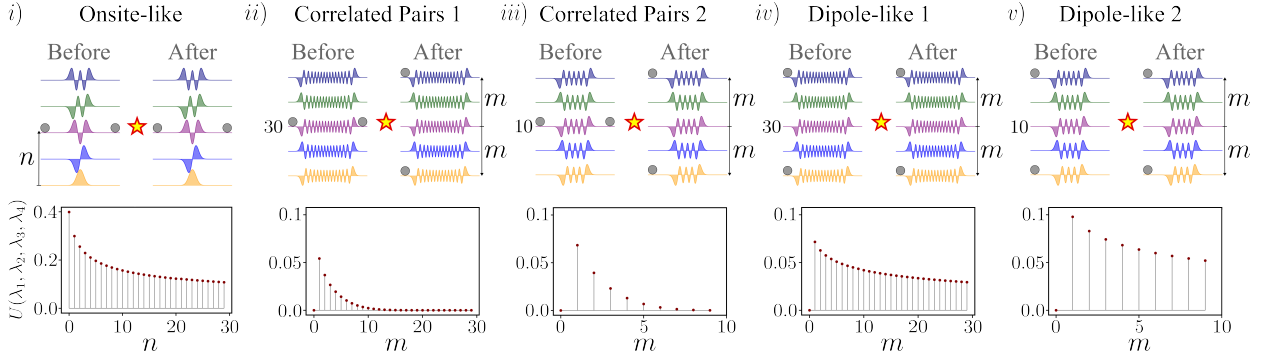}
\caption{\label{fig:interactiontypes} Examples of different types of interactions contained within Eq.~(\ref{Eq:rwaint}), including i) onsite, ii) + iii) correlated pair tunneling and iv) + v) dipole-like terms. For correlated pair and dipole-like interactions we present two cases for interactions centered around $\lambda = 30$ ii) and iv) and $\lambda=10$ iii) and v). In each case, the upper panel is a schematic with examples of where two atoms could be before and after the interaction. In the lower panel we then plot the corresponding relative interaction strength $U(\lambda_1, \lambda_2, \lambda_3, \lambda_4)$ as a function of either $n$ or $m$ as indicated in the upper panel. Note that no onsite-like terms are included within the correlated pair tunneling and dipole-like interactions to help separate out different effects.}
\end{figure*}
\begin{equation}
    h_{\lambda}(y) = \frac{1}{\sqrt{2^{\lambda}\lambda ! \sqrt{\pi}}}H_{\lambda}(y)e^{-\frac{y^2}{2}},
    \label{Eq:normalisedhermite}
\end{equation} 
with $H_{\lambda}(y)$ being the Hermite polynomial and where we have expressed $y$ in units of the harmonic oscillator length. Applying this basis transform to the mean-field contact interaction results in
\begin{eqnarray}
\mathcal{H}_{\text{int}} = \frac{g}{2l_H} \sum_{n, \lambda_1, \lambda_2, \lambda_3, \lambda_4} && a^{\dagger}_{n, \lambda_4}a^{\dagger}_{n, \lambda_3}a_{n, \lambda_2}a_{n, \lambda_1} \nonumber \\ && \times U(\lambda_1, \lambda_2; \lambda_3, \lambda_4),
\label{Eq:fullint}
\end{eqnarray}  where
\begin{equation}
    U(\lambda_1, \lambda_2; \lambda_3, \lambda_4) \equiv \int h_{\lambda_1}(y) h_{\lambda_2}(y) h_{\lambda_3}(y)h_{\lambda_4}(y) dy. 
\end{equation}
As discussed in the previous section, the interpretation of $\lambda$ as a synthetic dimension relies on a rotating wave approximation (RWA) in the high-frequency limit, which here requires $\omega_D\gg g$. To make this approximation, we map $a_{n,\lambda}$ into the rotating frame as 
\begin{equation}    
a_{n, \lambda}\equiv \psi_{n, \lambda}e^{i \omega_D \lambda t},
\end{equation}
and then neglect the fast oscillating terms to obtain~\cite{price2015}
\begin{eqnarray}
\mathcal{H}_{\text{int}} = \frac{g}{2l_H} \sum_{n, \lambda_1+\lambda_2 =\lambda_3 + \lambda_4} && \psi^{\dagger}_{n, \lambda_4}\psi^{\dagger}_{n, \lambda_3}\psi_{n, \lambda_2}\psi_{n, \lambda_1}\times \nonumber  \\ && U(\lambda_1, \lambda_2; \lambda_3, \lambda_4).
\label{Eq:rwaint}
\end{eqnarray}
Note, that the constraint introduced by this approximation, $\lambda_1 + \lambda_2 = \lambda_3 + \lambda_4$, enforces the conservation of energy. 
Combining this with the single-particle part of the Hamiltonian [Eq.~(\ref{Eq:HH})], we obtain the total effective Hamiltonian
\begin{equation}
\mathcal{H} = \mathcal{H}_{\text{HH}} + \mathcal{H}_{\text{int}},
\label{Eq:2ndquantmain}
\end{equation} 
which will be investigated numerically in the following sections.

To assist in the understanding of the above synthetic dimension interaction Hamiltonian [Eq.~(\ref{Eq:rwaint})], we can identify three prominent subcategories of interactions, inspired by the treatment in Ref.~\cite{Price2017}, which we call onsite, a ``correlated pair tunneling" and ``dipole-like" interactions. A schematic for each class of interactions is shown in the upper panels of Fig.~\ref{fig:interactiontypes} (i), (ii-iii) and (iv-v) respectively. (Note that these subsets do not exhaustively include all terms in the full RWA interactions above [c.f.~Eq.~(\ref{Eq:rwaint})].)

Firstly, onsite interactions correspond to those which occur when two particles are in the same atomic trap state before and after the collision, i.e. $\lambda_{1,2,3,4}\!=\!\lambda$ in Eq.~(\ref{Eq:rwaint}). The relative strength of these interactions is plotted in the lower panel of Fig.~\ref{fig:interactiontypes}~(i) as a function of $n\equiv \lambda$, the position along the synthetic dimension. These are the strongest type of interactions, but unlike usual contact interactions, they do decay algebraically along the synthetic dimension as the underlying trap states become increasingly delocalised. Note that in a real optical lattice, this would correspond to having a space-dependent lattice depth. 

Secondly, ``correlated pair tunneling" interactions correspond to pair-splitting or pair-forming terms in which two atoms either scatter away from or into the same trap state; for example, for pair-splitting these correspond to terms of the type  $\lambda_{1,2}=\lambda$, $\lambda_{3,4}\!= \lambda \pm m$ where $m$ is an integer in Eq.~(\ref{Eq:rwaint}), as shown schematically for two cases in the upper panel of Fig.~\ref{fig:interactiontypes} (ii-iii). We show two examples for $\lambda = 10$ and $30$ to highlight that the behavior of the relative interaction strength is consistent for different values of $\lambda$ for this type of interaction. The relative strength of such interactions is shown in the corresponding lower panels for the two examples as presented in ii) and iii). As can be seen, the strength of these processes is typically much less than that of the onsite terms and decays exponentially with increasing $m$. Note that we have excluded $m=0$ as this term corresponds to the onsite-like interaction plotted in panel (i). Interestingly, because the atoms have to scatter into/out of symmetrically distributed states, fewer of these processes are allowed as we approach either the lowest or highest $\lambda$ sites within a given system. This can be observed when comparing the possible interactions in Fig.~\ref{fig:interactiontypes} ii) and iii).

Finally, ``dipole-like" interactions correspond to those in which particles initially in states $(\lambda_1, \lambda_2)$ interact and either scatter to the same positions $(\lambda_1, \lambda_2)$ or equivalently (for indistinguishable bosons) flip-flop to states $(\lambda_2, \lambda_1)$. This is reminiscent of the effect of dipolar interactions. We can express these interactions as $\lambda_{1,2}=\lambda \pm m$, $\lambda_{3,4}\!= \lambda \pm m$, where $m$ is an integer  in Eq.~(\ref{Eq:rwaint}). As before the relative strength for two examples of these interactions is shown in the lower panels of Fig~\ref{fig:interactiontypes}(iv-v) for $\lambda=30$ and $\lambda=10$, respectively, as a function of $m$. Note that we have again excluded $m=0$, as this corresponds to the onsite-like interactions in panel (a). As can be seen, these interactions are generally stronger than the correlated pair tunneling terms and also only decay algebraically with increasing $m$.

For the purpose of our investigation we study the effects of weak interactions in the mean-field limit. In this regime the Gross-Pitaevskii equation effectively captures the relevant physics where the interactions are dominated by s-wave scattering.

\subsection{Numerical methods and observables} \label{sec:numerics}

We obtain the ground states of our system described by the extended Gross Pitaevskii equation by using the fourth order Runge Kutta method, and evolving in imaginary time until the differences in the wavefunction are $\sim\mathcal{O}(10^{-15})$. For all results, the system is normalized to $\sum_{n, \lambda} |\psi_{n,\lambda}|^2 = 1$.
As a reference, we check all our numerical calculations against known results for standard mean-field contact interactions, in which case the only non-zero interaction terms in Eq.~(\ref{Eq:2ndquantmain}) are given by $U(\lambda, \lambda, \lambda, \lambda)= 1$. 
We can also probe different subsets of the synthetic-dimension interactions by restricting the sum over $U(\lambda_1, \lambda_2, \lambda_3, \lambda_4)$ in Eq.~(\ref{Eq:2ndquantmain}) in our numerics.  

To analyze our results, we calculate the local density and current associated with each numerically-obtained wave function. The former is defined as usual by $|\psi_{n,\lambda}|^2$, while the latter is derived by considering
\begin{equation}
\dot{\rho}_{n, \lambda} = i \left[ \mathcal{H_{\text{HH}}}, \psi_{n,\lambda}^{\dagger}\psi_{n,\lambda} \right].
\end{equation} Here we define $\dot{\rho}_{n,\lambda}$ as the rate of change of the density operator, which is linked to the local current operator via the continuity equation. Note that here we neglect the contribution from $\mathcal{H_{\text{int}}}$ to $\dot{\rho}_{n,\lambda}$ as the resulting interaction contributions to the currents are much smaller than the kinetic energy contributions, both as we are in the weakly-interacting limit and because they consist of four-body terms, which are suppressed relative to the usual two-body terms.

From this approach, the corresponding local currents are then given by
\begin{equation}
J^{\uparrow}_{n, \lambda} =  t \left[i e^{i\varphi n}\psi^{\dagger}_{n, \lambda} \psi_{n, \lambda-1} + \text{h.c.}\right]
\end{equation}
along the synthetic dimension and  
\begin{equation}
J^{\rightarrow}_{n, \lambda} = J_x\left[i\psi^{\dagger}_{n, \lambda}\psi_{n-1,\lambda} + \text{h.c.}\right]  
\end{equation}
along the real spatial direction.
\begin{figure*}
\includegraphics[width=\textwidth]{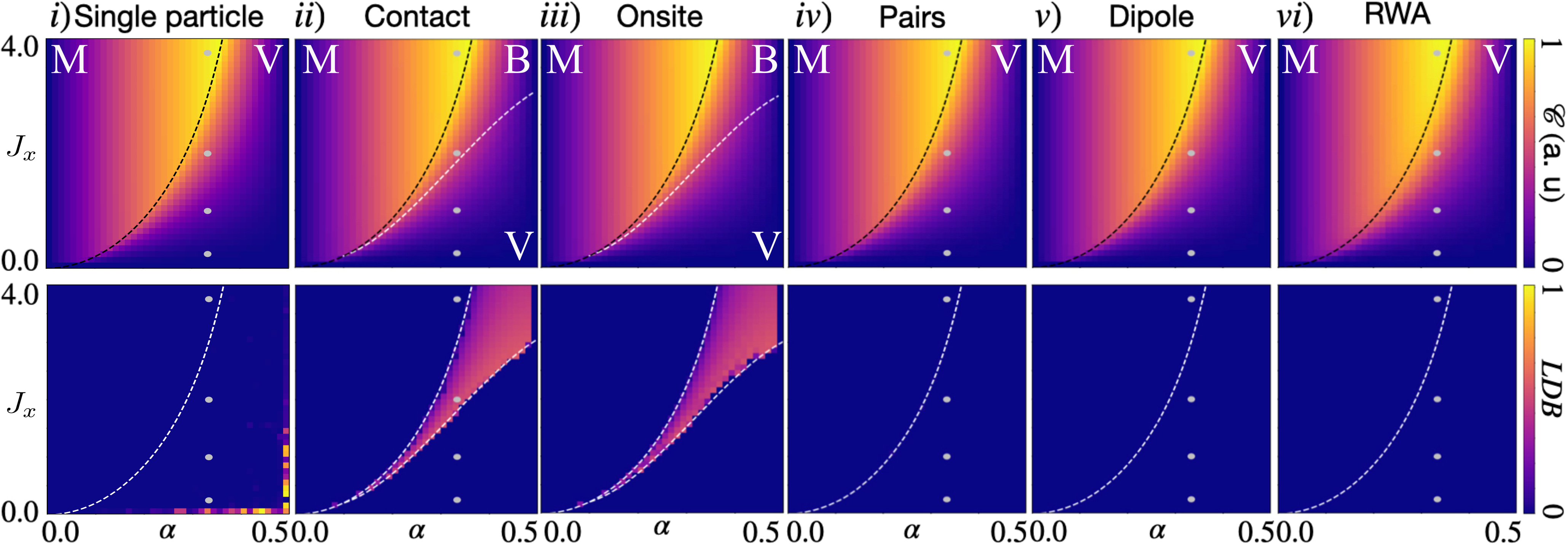}
\caption{\label{fig:phasediagram} The chiral current [Eq.~(\ref{eq:chiral})] (upper panels) and ladder density bias (LDB) [Eq.~(\ref{eq:ldb})] (lower panels) as a function of magnetic flux $\alpha$ and inter-leg hopping $J_x$, calculated for the imaginary-time-evolved state of a $59$-site ladder with (i) no interactions, (ii) contact, (iii) onsite, (iv) correlated pair tunneling, (v) dipole-like and (vi) full RWA synthetic-dimension interactions, with $g=0.5$ and $t=1$. Silver dots indicate those example states plotted in Figs.~\ref{fig:laddersingleparticle}-\ref{fig:ladderrwa} and in Appendix B. Labels (M), (B) and (V) indicate when the ground-state is in the Meissner, LDB and vortex phase respectively.  Note that we have normalized the chiral current to one for presentation purposes. The (upper) dotted line in each panel indicates the critical anisotropy [Eq.~(\ref{eq:critical})]; this is in good agreement with the maximum saturation of the chiral current as expected~\cite{Atala_2014}. As can be seen, the chiral current does not appear to be significantly affected by the choice of interactions; however, the LDB does change, with a non-zero ladder density bias appearing for contact and on-site synthetic dimension interactions reflecting the presence of the biased ladder phase~\cite{wei2014theory}. In these cases, the lower dotted line is a guide to the eye, indicating the boundary between this and the vortex-lattice phase. Note that the non-zero LDB in the single-particle case (around $J_x\approx 0$ and $\alpha \approx 0.5$) is due to numerical artifacts.}
\end{figure*}
For the HH ladder in Sec.~\ref{sec:ladder}, we will also calculate two further observables; firstly, the density bias between the two legs of the ladder, defined as
\begin{equation}
LDB\equiv \mid \mathcal{D}_2 - \mathcal{D}_1 \mid,
\label{eq:ldb}
\end{equation} 
where the total density along each ladder leg  is given by
\begin{equation} \label{eq:density}
\mathcal{D}_{l}= \sum_\lambda \psi^{\dagger}_{l, \lambda}\psi_{l, \lambda},
\end{equation} where $l\!=\!1,2$ denotes the ladder leg. Secondly, we will calculate the total chiral current  
\begin{equation}
\mathcal{C} = \mid\mathcal{J}_1 - \mathcal{J}_2 + \mathcal{J}_{\text{edge}} \mid , \label{eq:chiral}
\end{equation}
where the first two terms correspond to the average local current along each ladder leg
\begin{equation} \label{eq:current}
\mathcal{J}_{l} = \frac{1}{N_\lambda}\sum_{\lambda}^{N_{\lambda}} J^{\uparrow}_{l, \lambda},
\end{equation}
while the last term corresponds to current contributions from the top and bottom of the finite ladder, 
\begin{equation} 
\mathcal{J}_{\text{edge}} = \frac{1}{2} \left[ \mathcal{J}^{\rightarrow}_{1, (N_\lambda-1)} - \mathcal{J}^{\rightarrow}_{1, 0} \right]. 
\end{equation}
Note that in practice, this contribution can also be neglected as the open boundary conditions along the ladder cause the density to fall towards zero at these edges [c.f. Fig.~\ref{fig:laddersingleparticle}]. 
Together, the chiral current and ladder density bias will allow us to characterize the phase diagram of the HH ladder with interactions, as we shall now discuss.

\section{HH Ladder With Synthetic-Dimension Interactions}
\label{sec:ladder}

We shall begin by reviewing the physics of the two-leg HH ladder in the single-particle regime [Section~\ref{sec:single_particle}] and  with contact interactions [Section~\ref{Sec:twolegladdercontact}]. In both cases it is known that ladders exhibit a rich phase diagram as a function of the magnetic flux $\alpha$ and the hopping anisotropy $J_x/t$~\cite{wei2014theory, Atala_2014,piraud2015vortex, kelecs2015mott,qiao2021quantum,uchino2015population,natu2015bosons,kolovsky2017bogoliubov, impertro2025strongly}. In Section~\ref{sec:laddersynthetic} we investigate the effect that introducing the harmonic synthetic interactions along the ladder direction has on the previously understood phases.
Note that while we shall focus  on hybrid HH ladders with a synthetic dimension of atomic trap states, previous works have also investigated the effects of interactions on other hybrid real-synthetic variants of the two- and three-leg HH ladder, for example, where each leg corresponds to a different hyperfine state with interactions that are SU(N)-invariant~\cite{uchino2016analytical, petrescu2015chiral, greschner2016symmetry,buser2020interacting}. 

\subsection{Single-Particle Physics}
\label{sec:single_particle}
In the single-particle limit, the two-leg HH ladder Hamiltonian [Eq.~(\ref{Eq:HH})] can be straightforwardly diagonalised to find two single-particle bands described by~\cite{wei2014theory, Atala_2014,kelecs2015mott,uchino2015population}
\begin{equation}
 E_{\pm} (k) = - 2 t \cos k \cos \pi \alpha\pm\sqrt{4 t^2 \sin^2 k \sin^2 \pi \alpha  + J_x^2} ,
\end{equation}
where $k$ is the momentum along the ladder-leg. Above a critical anisotropy,
\begin{equation}
J_x^c = 2 t \tan \pi \alpha \sin\pi\alpha,\label{eq:critical}
\end{equation}
the lower band has a single minimum at $k=0$, while otherwise, the lower band has two minima at non-zero momenta $k=\pm k_0$. 

The ground state therefore changes as a function of flux and hopping anisotropy between the single $k\!=\!0$ state and a superposition of the two $\pm k_0$ states. The former corresponds to the Meissner phase, so-called because in real space, this state has uniform chiral currents flowing in opposite directions along each leg (with negligible rung currents), mimicking the Meissner effect in superconductivity [c.f. Fig.~\ref{fig:laddersingleparticle} below].  The latter corresponds to a vortex-lattice phase, so-called because currents circulate around plaquettes of the ladder in a manner reminiscent of vortices. The density is also spatially-varying in this phase, for which reason it is sometimes referred to as the modulated density phase~\cite{wei2014theory}.

The phase transition between the Meissner and vortex phases can be visualised by plotting the value of the chiral current [Eq.~(\ref{eq:chiral})] as a function of magnetic flux and anisotropy as shown in the upper panel of Fig.~\ref{fig:phasediagram}(i) for a single-particle two-leg ladder with 59 sites. Here, the dashed black line corresponds to the analytical prediction for the critical anisotropy $J_x^c/t$ (note that $t=1$ in these numerics). As can be seen, this curve is in excellent agreement with the maximum of the observed chiral current. For completeness, we have also plotted the ladder density bias [Eq.~(\ref{eq:ldb})] in the lower panel of Fig.~\ref{fig:phasediagram}(i); however, this vanishes in both the Meissner and vortex-lattice phases. 

When the system is in the Meissner phase [i.e. above the critical line in Fig.~\ref{fig:phasediagram}(i)], the chiral current is maximal and fully screens the magnetic field; thus removing the periodicity enforced by the magnetic length. In an infinite system, the maximal chiral current is given by $\mathcal{C}_{\text{max}} = 2 t \sin \pi \alpha$~\cite{wei2014theory}, which is independent of $J_x$ but increasing with magnetic flux. For this reason, this phase has sometimes been referred to in the literature as the saturated chiral current phase~\cite{wei2014theory}. An example of this phase is shown for a numerically-obtained ground state in Fig.~\ref{fig:laddersingleparticle}(iv) with $N_\lambda=59$, for $\alpha\!=\!1/3$ and $J_x\!=\!3.75$ [i.e. parameters corresponding to the uppermost silver dot in Fig.~\ref{fig:phasediagram}(i)]. Here (and in all following figures), the density is represented by the size and color of the dots, while the current is depicted by the magnitude and direction of the arrows drawn connecting the dots (as shown in the zoomed-in inset for each panel). Note that both the maximum local density and current are normalized for each panel for clarity of presentation. As can be seen, the Meissner phase exhibits, as expected, strong currents flowing in opposite directions along the two legs, with 
very small currents flowing in between. There is also an overall density modulation along the ladder, due to the finite system size.  

When the system is in the vortex-lattice phase [i.e. below the critical line in Fig.~\ref{fig:phasediagram}(i)], the chiral current begins to decrease as the magnetic field starts to penetrate the system. This behavior is analogous to that of a superconductor undergoing a transition from the Meissner effect to an Abrikosov vortex lattice. Three examples of this phase are given in Fig.~\ref{fig:laddersingleparticle} for $\alpha\!=\!1/3$ and (i) $J_x\!=\!0.25$, (ii) $J_x\!=\!1$ and (iii) $J_x\!=\!2$ [i.e. parameters corresponding to the lower three silver dots in Fig.~\ref{fig:phasediagram}(i) respectively]. As can be seen, these phases are characterized by plaquettes with circulating currents, similar to vortices. The length of these plaquettes increases with increasing $J_x$; this can be understood in terms of the positions, $\pm k_0$, of the single-particle minima shifting towards zero as the phase transition is approached~\cite{Atala_2014}. 

\subsection{Standard Mean-field Contact Interactions}
\label{Sec:twolegladdercontact}
\begin{figure}
    \centering
    \includegraphics[width=\linewidth]{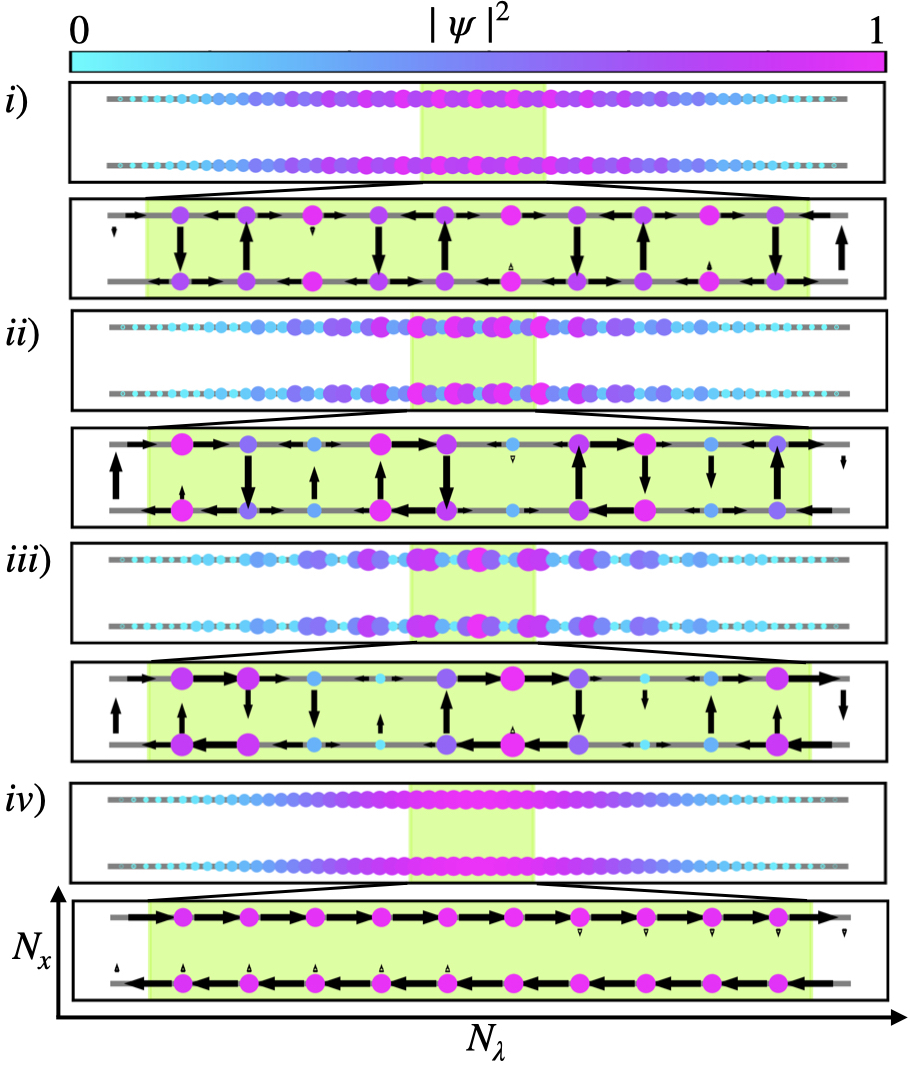}
    \caption{Example single-particle ground states  obtained numerically for $\alpha\!=\!1/3$, $t\!=\!1$ and $N_\lambda=59$ with (i) $J_x\!=\!0.25$, (ii) $J_x\!=\!1$, (iii) $J_x\!=\!2$, (iv) $J_x\!=\!3.75$,
    i.e. parameters corresponding to the silver dots marked in Fig.~\ref{fig:phasediagram}(i). The density (current) is represented by the size and color (direction) of the dots (arrows). Note that the local density and current are both normalized for each panel to unity. The upper portion of each panel shows the density for the entire system, while the lower portion is a zoomed-in inset of the density and current for the section highlighted in green. As can be seen, panels (i)-(iii) are in the vortex-lattice phase with  circulating plaquettes, while panel (iv) is in the Meissner-phase with a strong chiral current.}
    \label{fig:laddersingleparticle} \label{sec:laddercontact}
\end{figure}

\begin{figure}
    \centering
    \includegraphics[width=\linewidth]{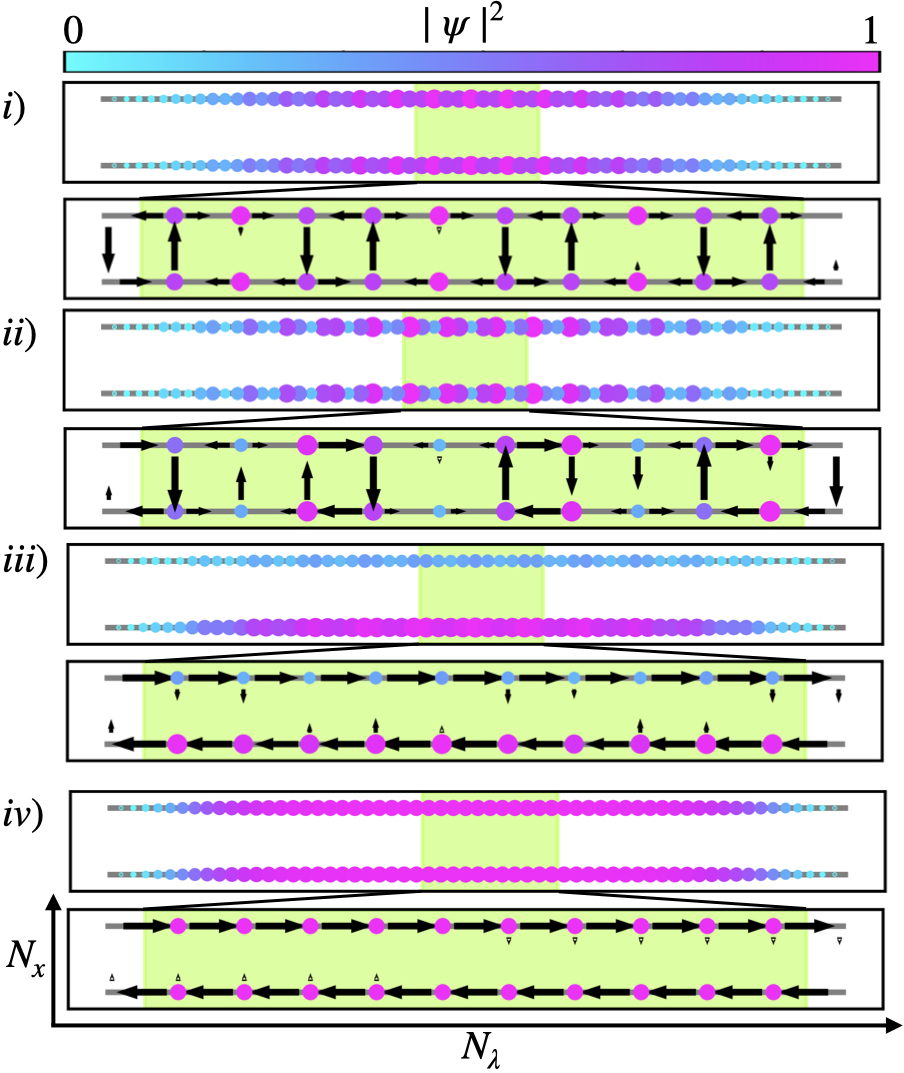}
    \caption{Example ground states obtained numerically for $\alpha\!=\!1/3$, $t\!=\!1$, $N_\lambda \!=\!59$ and standard contact interactions of strength $g\!=\!0.5$, with (i) $J_x\!=\!0.25$, (ii) $J_x\!=\!1$, (iii) $k\!=\!2$, (iv) $J_x\!=\!3.75$,
    i.e. parameters corresponding to the silver dots marked in Fig.~\ref{fig:phasediagram}(ii). Most features closely resemble those of the single-particle states in Fig.~\ref{fig:laddersingleparticle}, except for the appearance of the biased ladder phase in panel (iii) and a broader overall density distribution in all cases due to interactions.}
    \label{fig:laddercontact}
\end{figure}

Introducing mean-field contact interactions leads to the numerical phase diagrams shown in Fig.~\ref{fig:phasediagram}(ii), where, as before, the upper dotted line indicates the single-particle critical anisotropy $J_x^c/t$. Note that although interactions do shift this critical line, the effect is small here as the interactions are weak; hence, the single-particle line is still in good agreement with the observed maximum in the chiral current. 

The region above the critical line is also unchanged compared to the single-particle case; this is because for weak interactions, the Meissner phase is preserved~\cite{wei2014theory}. A numerical example of this state is shown in Fig.~\ref{fig:laddercontact}(iv) for $J_x\!=\!3.75$ and $\alpha=1/3$ [i.e. parameters corresponding to the uppermost silver dot in Fig.~\ref{fig:phasediagram}(ii)], with $t=1$, $g=0.5$ and $N_\lambda=59$. As can be seen, this state strongly resembles the single-particle Meissner phase shown in Fig.~\ref{fig:laddersingleparticle}(iv); however, the overall density envelope along the ladder is different, as the interactions cause the wave-function to spread out, leading to a Thomas-Fermi-like rather than a sinusoidal free-particle profile. 

Below the critical line, there are also now two distinct phases as can be clearly identified from the ladder density bias [Eq.~(\ref{eq:ldb})], as shown in the lower panel of Fig.~\ref{fig:phasediagram}(ii). In particular, immediately below the Meissner phase, the LDB takes a non-zero value; this is often referred to as the ``biased ladder phase"~\cite{wei2014theory}. An example of this state is shown in Fig.~\ref{fig:laddercontact}(iii) for $J_x\!=\!2$ and all other parameters as above; here it can be seen that there are still strong Meissner-like currents, but now with a significant density imbalance between the two legs. This phase can be captured, for example, through a variational approach, which predicts that all bosons condense into one of the two minima of the single-particle bandstructure~\cite{wei2014theory}. Note that this quantity dictates our choice of $g$. Specifically, the value of the ladder bias varies significantly with respect to $g$ for values of $g < 0.5$ before plateauing for $g \approx 0.5$. Note also, that this phase is two-fold degenerate as the choice of which leg has higher density is arbitrary. In our numerics, the specific states obtained, e.g. in Fig.~\ref{fig:laddercontact}, depends on the random initial state used for the imaginary-time evolution [c.f. Sec.~\ref{sec:numerics}]. Note that selecting random initial states allow us to identify symmetry broken ground-states where relevant.

Even further below the critical line, there is a second phase transition, after which the LDB drops again to zero. This is again the vortex-lattice phase in which currents circulate around plaquettes within the ladder, while the local density is modulated. Examples of this phase are shown in Fig.~\ref{fig:laddercontact}(i)  and (ii) for $J_x\!=\!0.25$ and $J_x\!=\!1$ respectively. This phase can also be described by the same variational approach above, which now predicts that the bosons occupy both single-particle minima~\cite{wei2014theory}. As a result, these phases again strongly resemble those in the single-particle case [c.f. Fig.~\ref{fig:laddersingleparticle}(i) and (ii)], up to differences in the overall density modulation.

\subsection{Synthetic-Dimension Interactions}\label{sec:laddersynthetic}

Having reproduced the previously-known results, we now include the new form of interactions along the synthetic dimension of harmonic trap states [Eq.~(\ref{Eq:2ndquantmain})]. Before proceeding to the full RWA terms, we briefly discuss the different subsets of onsite, correlated pair tunneling and dipole-like interactions identified previously in Sec.~\ref{sec:inter}. 

\begin{figure}
    \centering
    \includegraphics[width=\linewidth]{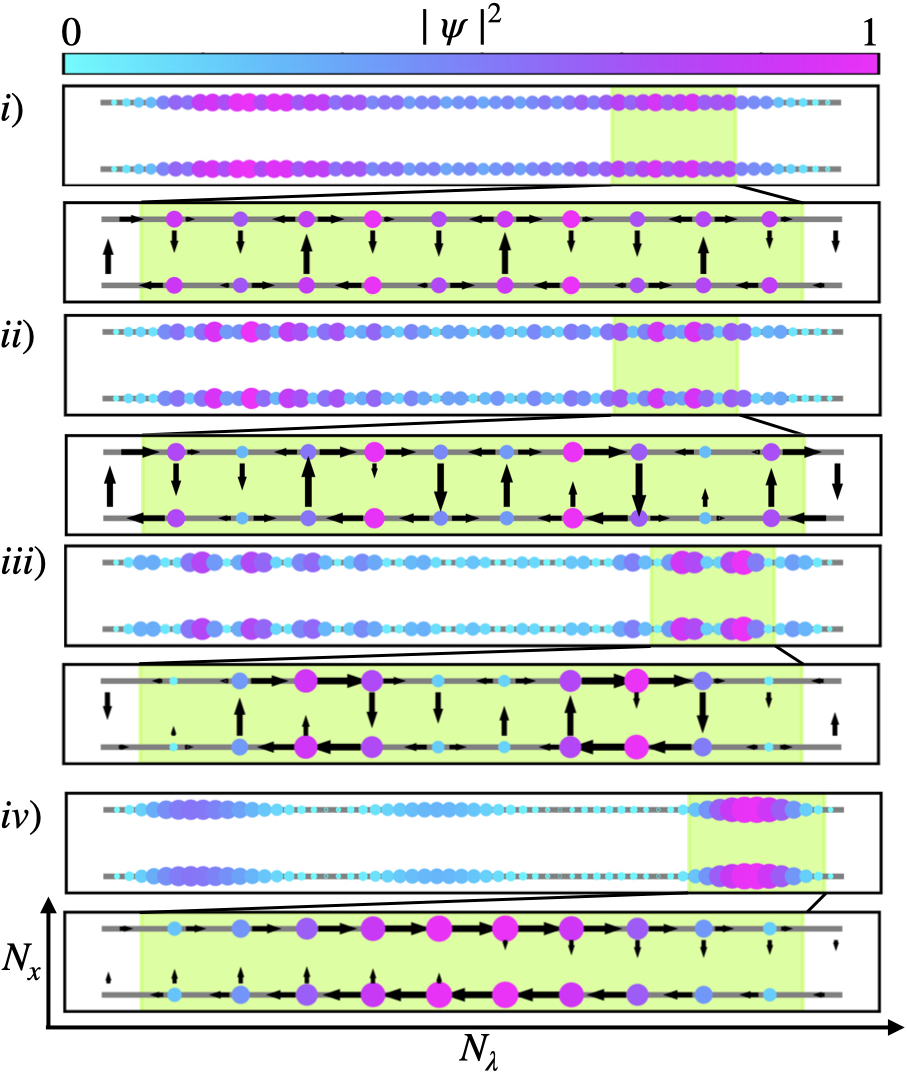}
    \caption{Example ground states obtained numerically for $\alpha\!=\!1/3$, $t\!=\!1$, $N_\lambda\!=\!59$ and full RWA synthetic-dimension  interactions, with (i) $J_x\!=\!0.25$, (ii) $J_x\!=\!1$, (iii) $J_x\!=\!2$, (iv) $J_x\!=\!3.75$,
    i.e. parameters corresponding to the silver dots marked in Fig.~\ref{fig:phasediagram}(vi). As can be seen, panels (i)-(iii) are variants of vortex-lattice phases, without the biased ladder phase [c.f Fig.~\ref{fig:laddercontact}], while panel (iv) is a type of Meissner phase. The inhomogeneity of the interactions has led to overall density modulations, including a build-up of particles at either end of the synthetic dimension.}
    \label{fig:ladderrwa}
\end{figure}

Firstly, for onsite interactions, we obtain qualitatively similar results to those for standard contact interactions described above. This is evident, for example, from the phase diagrams shown in Fig.~\ref{fig:phasediagram}(iii). This is to be expected, as the only difference between these cases is that the synthetic-dimension onsite interactions decay with position along the synthetic dimension, while the standard contact ones do not. However, since this decay is relatively slow, the strength of the on-site interactions is approximately uniform over the majority of the ladder [cf. Fig.~\ref{fig:interactiontypes}] and so no significant differences in the ground state are seen. 

Secondly, for correlated pair tunneling and dipole-like interactions, the main difference that we observe compared to the previous results is the disappearance of the biased density ladder phase; this can be seen in Fig.~\ref{fig:phasediagram}(iv) and (v), which again resemble that of the single-particle case. Instead of the biased ladder phase, the numerically-obtained ground states for the correlated pair tunneling and dipole-like interactions are again variants of the vortex lattice phase, with circulating plaquettes; examples of these are shown in Fig.~\ref{fig:ladderelastic}-\ref{fig:ladderdipole} in Appendix B. Note that for the dipole-like case, the peak density also becomes shifted towards higher $\lambda$ states by an amount that increases with increasing $J_x$; this is perhaps because the overall magnitude of the dipole-like interactions decreases with position up the synthetic dimension. 

Finally, we include the full RWA synthetic-dimension interactions. As can be seen in Fig.~\ref{fig:phasediagram}(vi), the phase diagram looks qualitatively identical to that for the previous 
cases, with a vanishing ladder density bias. This suggests that the underlying physics is similar to that observed in normal Hofstadter ladders. Indeed, specific examples of the numerically-obtained ground states are shown in Fig.~\ref{fig:ladderrwa} for the same parameters as in previous figures; as can be seen, these are variants of vortex-lattice and Meissner phases, except now with an unusual density modulation such that there are large density peaks towards either end of the synthetic dimension, reflecting the inhomogeneity of the interaction terms. As $J_x$ increases further, this density modulation also increases in strength, leading to a further build-up of particles near the synthetic-dimension boundaries. Interestingly, therefore, adding interactions to a synthetic dimension of harmonic trap states appears to preserve the small-scale structure of vortex-lattice and Meissner phases in a Hofstadter ladder, while removing the biased ladder phase found for usual contact interactions.  

\section{2D Harper-Hofstadter Lattice with Synthetic-Dimension Interactions}
\label{sec:vortex_ground_states}

 We now extend our discussion to the mean-field 2D Harper-Hofstadter model where one direction corresponds to real space and the other to a synthetic dimension of atomic trap states. To provide a baseline for our results, we begin in Section~\ref{sec:HHcontact} by revisiting the case of a 2D real-space HH model with standard contact interactions in both directions. We then proceed to study the effects of introducing synthetic-dimension interactions in Section~\ref{sec:third}.  

\begin{figure}
\centering
\includegraphics[width=\columnwidth]{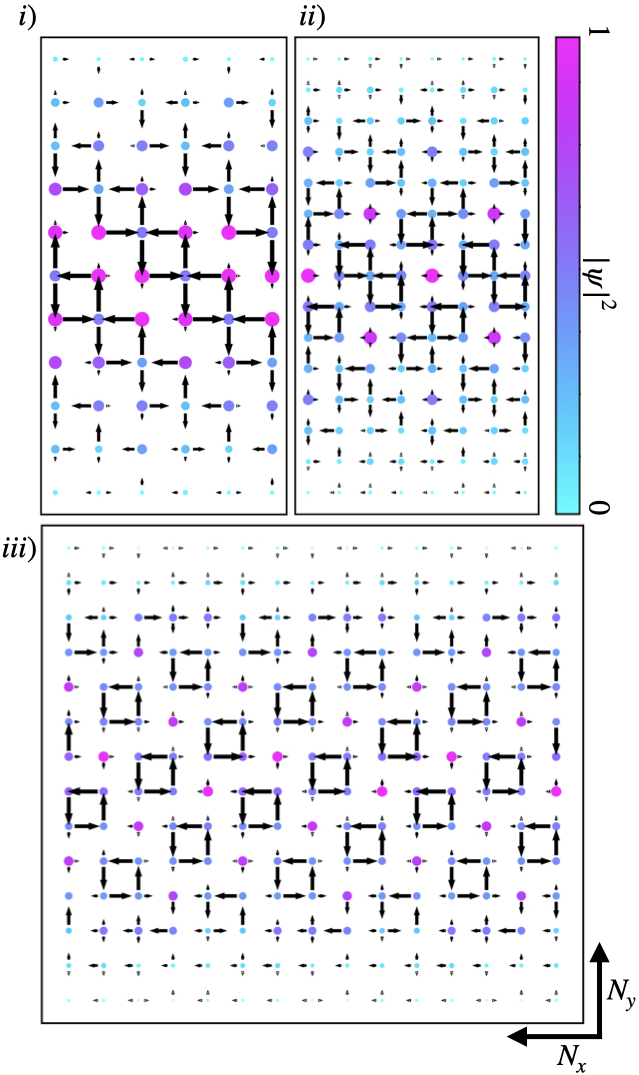}
\caption{\label{fig:standardcontactinteractions} Ground states of the HH model with standard mean-field contact interactions with $t\!=\!J_x\!=\!1$ and $g\!=\!0.1$ and (i)~$\alpha\! =\! 1/3$, (ii)~$\alpha\! =\! 1/4$, (iii)~$\alpha\! =\! 1/5$. The system sizes are respectively $(N_x, N_y) \!=\! (6, 11), (8,15), (15, 14)$, with periodic (open) boundary conditions along x (y). As in Section~\ref{sec:ladder}, the size and color of the dots are proportional to the local density, while the black arrows indicate the size and direction of the nearest-neighbor local currents (the length-scale of the arrows is not maintained between panels). The configurations shown are each one of a discrete set of degenerate states related by the broken symmetries of the system; these results are consistent with those found for the case of periodic boundary conditions studied in Ref.~\cite{PhysRevA.83.013612}. }
\end{figure}

 \subsection{Standard mean-field contact interactions} \label{sec:HHcontact}

The 2D real-space Harper-Hofstadter model with contact interactions has previously been studied, for example, in Ref.~\cite{PhysRevA.83.013612}, which employed a wave-function ansatz based on Gutzwiller mean-field theory~\cite{PhysRevLett.81.3108, PhysRevA.77.033629} in the weak-interaction limit. For magnetic fluxes of the form $\alpha=p/q$, the condensate then exhibits a $q \times q$ unit cell in real space, with a reduced spatial symmetry compared to the underlying square lattice. Similar to the vortex-lattice phase in the Harper-Hofstadter ladder [c.f. Sec.~\ref{sec:ladder}], these states have highly non-uniform condensate densities, with large local currents circulating  counter-clockwise around certain plaquettes of the underlying lattice, similar to how current circulates around a vortex core. As $\alpha$ decreases, the density of these circulating plaquettes decreases until for larger values of $q$, i.e. $q=5$, the states resemble Abrikosov vortex lattices~\cite{PhysRevA.83.013612}.

Due to the inhomogeneity of our synthetic-dimension interactions [c.f. Sec.~\ref{sec:inter}], we are restricted to studying systems with a finite synthetic dimension with open boundary conditions. To provide a baseline for our study, we therefore start by numerically reproducing the results of Ref.~\cite{PhysRevA.83.013612} but now using the imaginary-time evolution method [c.f. Sec.~\ref{Sec:scheme}] for a system with open (periodic) boundary conditions along the y (x) direction. The results of these calculations are shown in Fig.~\ref{fig:standardcontactinteractions} for $t\!=\!J_x\!=\!1$ and $g\!=\!0.1$ and (i)~$\alpha\! =\! 1/3$, (ii)~$\alpha\! =\! 1/4$, (iii)~$\alpha\! =\! 1/5$. Note that our choice of $g$ is arbitrary but the states observed are consistent for a range of values [see Appendix~C]. The system sizes are chosen respectively as  $(N_x, N_y) \!=\! (6, 11), (8,15), (15, 14)$, so that $N_x$ is  a multiple of $q$ to ensure commensurability in the periodic direction. Due to the open boundary conditions along $\lambda$ the observed commensurability condition is a multiple of $q$, minus one. In cases where non-commensurate values are chosen we do not reproduce the known results from Ref.~\cite{PhysRevA.83.013612}. For sufficiently large $N_{y}$, differences between states with commensurate and non-commensurate $N_{y}$ sites become negligible. As expected, the obtained states each exhibit a distinctive $q \times q$ unit cell near the center of the system in the y direction, where the density is primarily localized. These observed configurations are in very good qualitative agreement with the infinite system results of Ref.~\cite{PhysRevA.83.013612}.  The main difference in our case is that the condensate density drops towards zero at the edges along the y direction due to the open boundary conditions; for the weak interactions chosen here, this fall-off can be captured approximately by a sinusoidal envelope.

We note that, as the states in Fig.~\ref{fig:standardcontactinteractions} spontaneously break certain symmetries of the underlying square lattice, each of these configurations is only one of a discrete set of degenerate states. These different states are equivalent under translations or rotations and/or reflections; see Ref.~\cite{PhysRevA.83.013612} for a full enumeration of the degeneracies and associated symmetries for these chosen values of the magnetic flux. As in the ladder case the specific state obtained is dependent on the initial random state chosen [c.f. Sec.~\ref{sec:numerics}].

\begin{figure}
    \centering    \includegraphics[width=\linewidth]{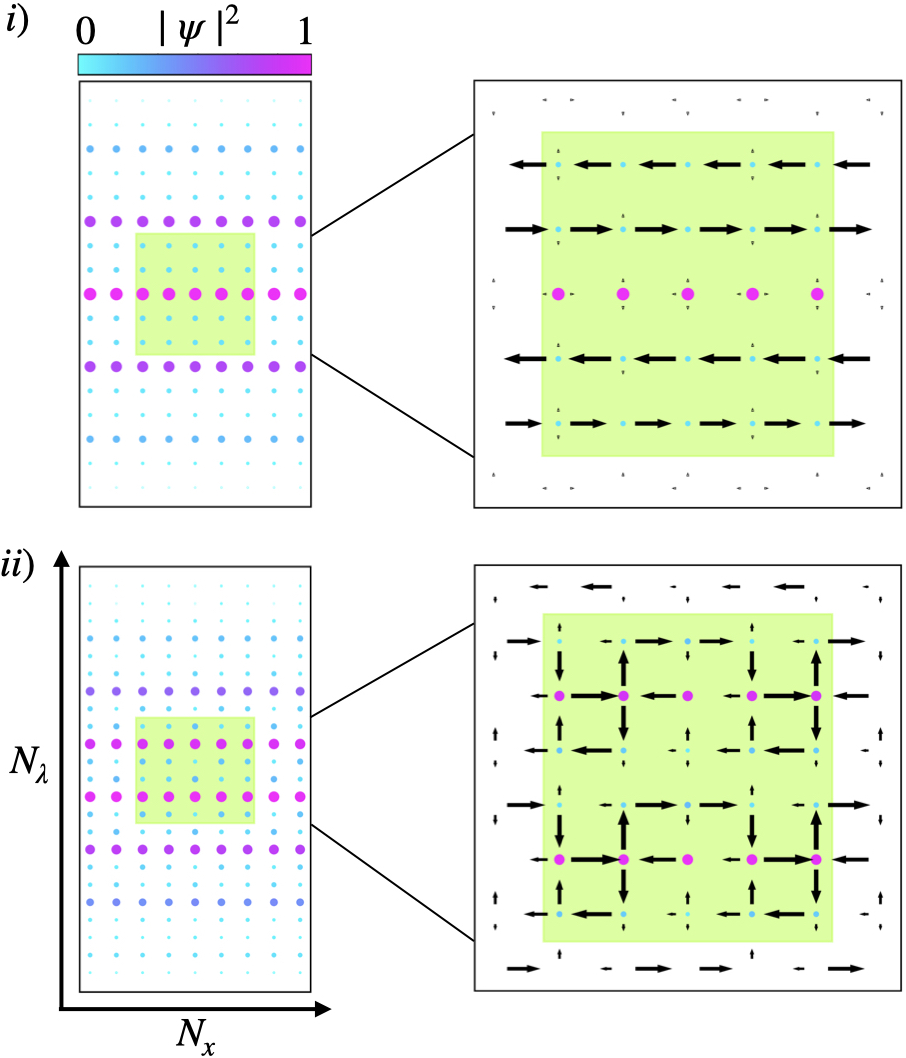}
    \caption{Ground states of the HH model with RWA synthetic-dimension interactions for $t\!=\!1$, $J_x\!=\!1$, $g\!=\!0.1$, $N_x\!=\!9$ and $\alpha \!= \!1/3$ for (i) $N_\lambda \!=\! 17$ and (ii) $N_\lambda \!= \!23$. (Note that the latter is one of two states that are energetically degenerate within numerical accuracy at this system size, with the other being a Meissner stripe state similar to that shown in panel (i)).) For each panel, the left-hand part shows the local density over the entire system, while the right-hand part is a zoomed-in inset of the density and current for the example area highlighted in green.}
    \label{fig:hhmodelrwa1d3}
\end{figure}

\subsection{Synthetic-dimension interactions} \label{sec:third}
Focusing firstly on the case of $\alpha\!=\!1/3$, we consider a 2D HH model in the $x-\lambda$ plane with the full RWA synthetic-dimension interactions introduced in Section~\ref{sec:inter}. The corresponding numerically-obtained configurations are shown in Fig.~\ref{fig:hhmodelrwa1d3} for $t\!=\!J_x\!=\!1$ and $g\!=\!0.1$, and for the examples of (i) $N_\lambda = 17$ and (ii) $N_\lambda = 23$. Note that throughout we have again applied periodic boundary conditions along the real $x$ direction; we have checked that for open boundary conditions, the same types of states are found, albeit with an overall density envelope. Interestingly, the inclusion of synthetic-dimension interactions dramatically changes the structure of the condensate state from that seen for contact interactions [c.f. Fig.~\ref{fig:standardcontactinteractions}(i)]. A key feature of these new states is the appearance of a strong local density modulation along the synthetic dimension with a period of 3 sites, reflecting the periodicity associated with the magnetic flux $\alpha\!=\!1/3$, even while the density along the real dimension is (approximately) uniform. 

\begin{figure}
    \centering    \includegraphics[width=\linewidth]{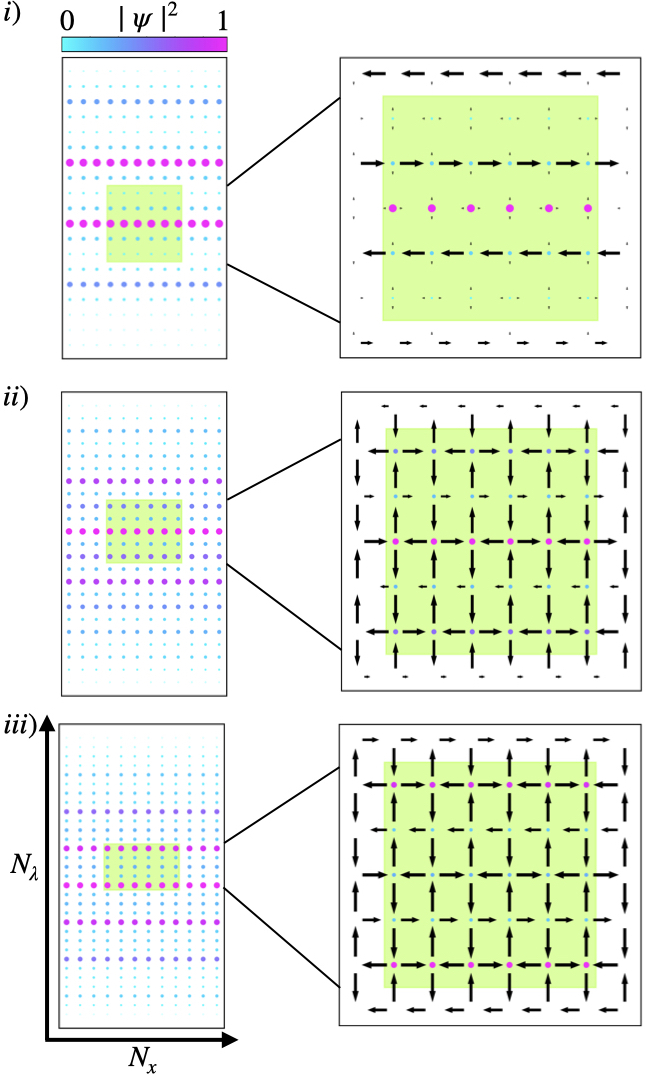}
    \caption{Ground states of the HH model with RWA synthetic-dimension interactions for $\alpha\!=\!1/4$ and $N_x\!=\! 12$, and for i) $N_\lambda \!= \!19$, ii) $N_{\lambda}\! = \!23$ and iii) $N_\lambda\!=\!31$, with all other parameters as in Fig.~\ref{fig:hhmodelrwa1d3}. As $N_\lambda$ increases, the ground state changes from a Meissner stripe state to one with arrays of plaquettes, with strong alternating currents.}
    \label{fig:hhmodelrwa1d4}
\end{figure}

\begin{figure}
    \centering    \includegraphics[width=\linewidth]{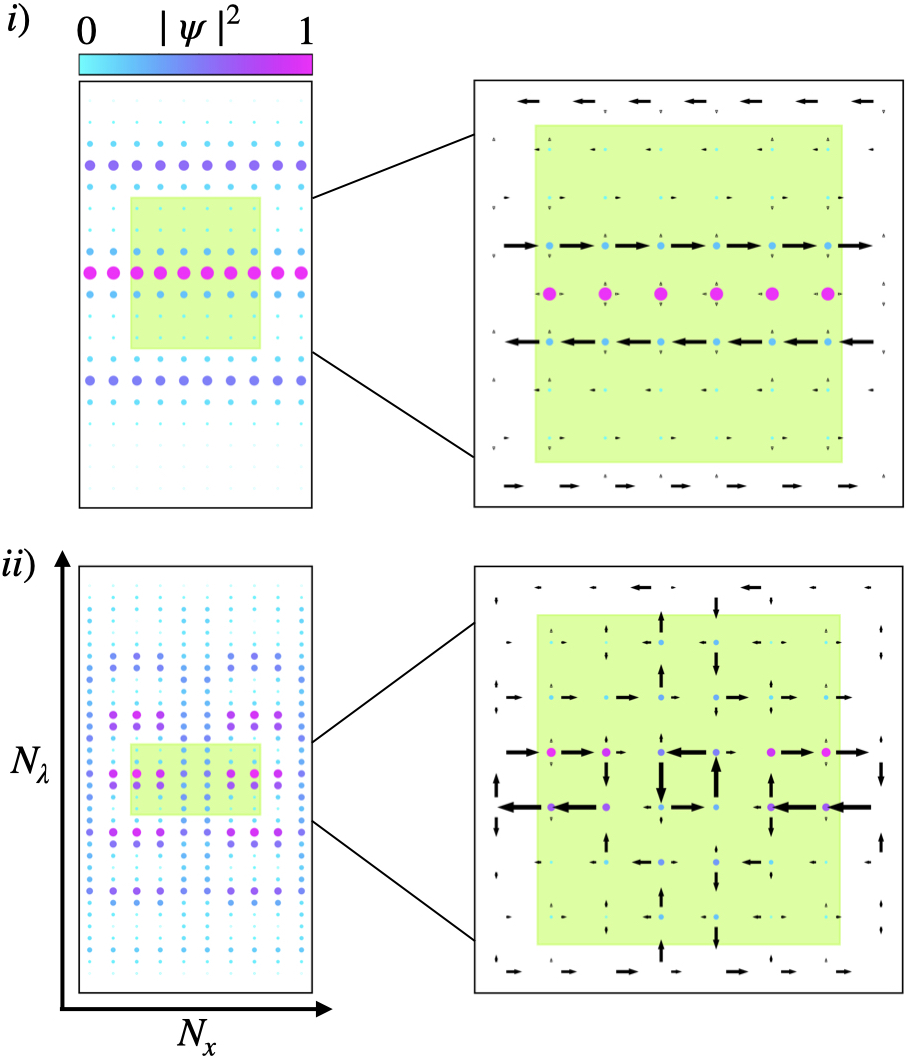}
    \caption{Ground states of the HH model with RWA synthetic-dimension interactions for $\alpha\!=\!1/5$, $N_x\!=\!10$ and $g\!=\!0.1$, and for i) $N_{\lambda}\!=\!19$  and ii) $N_{\lambda}\!=\!34$ with all other parameters are above. As $N_\lambda$ increases, the ground state changes from a Meissner stripe state to one with strong density modulations in both directions and currents circulating around different plaquettes.}
    \label{fig:hhmodelrwa1d5}
\end{figure}

For the smaller system size in (i), the ground state is a Meissner-like state, albeit with currents that flow along the real spatial dimension instead of along the synthetic dimension [c.f. Sec.~\ref{sec:ladder}]. As can be seen, this state is characterized by a row of high-density sites with negligible currents, alternating with rows of low-density sites hosting equal and opposite currents; we henceforth call this a ``Meissner stripe" state. We have found, by analyzing different subsets of the synthetic-dimension interactions, that including only dipole-like interactions again leads to a Meissner stripe state, while having either onsite or correlated pair tunneling interactions instead reproduces the standard contact-interaction vortex-lattice case (see Appendix B). This suggests that dipole-like interactions play a key role in determining the ground state structure of the condensate with the full RWA synthetic-dimension interactions for these parameters. 

Interestingly, for the particular system size of $N_\lambda = 23$, 
we find both a Meissner stripe state but also the state shown in panel (ii), which is energetically degenerate with the former to within the accuracy of our calculations, but which instead has some identifiable circulating plaquettes. This is reminiscent of the ``vortex lattice" found for standard contact interactions [c.f. Fig.~\ref{fig:standardcontactinteractions}(i)], although here the largest current circulation is clockwise rather than counter-clockwise. This state is not found for the larger system sizes we tested, so further study would be necessary to discern if it is a true ground state of the system. However, similar states can sometimes be found as the sole ground state by ramping up the interaction strength [see Appendix C]. This suggests that while the Meissner stripe state usually dominates for $\alpha=1/3$, it is possible to tune system parameters to realize a more vortex-lattice-like state.  

Turning next to $\alpha=1/4$, we find ground states such as those shown in Fig.~\ref{fig:hhmodelrwa1d4} for (i) $N_\lambda = 19$, (ii) $N_{\lambda}\! =\! 23$ and (iii) $N_\lambda=31$, with other parameters as above. Similar to the case of $\alpha=1/3$, the ground-state density appears to be uniform  along the real dimension, but strongly modulated along the synthetic dimension, albeit now with a period of 4 sites, reflecting the larger magnetic unit cell. For a small system size as in panel (i), we again find a type of Meissner stripe state, with currents flowing along the real dimensions. The main difference between this state and that observed for $\alpha=1/3$ is the appearance of an extra very low density row of sites between the current-carrying rows, as required by the larger magnetic unit cell. By analysing different subsets of interactions separately, we again find that the same Meissner stripe state appears with dipole-like interactions, suggesting that these interactions are dominating [see Appendix B]. 

However, we also observe that, as the system's length along the synthetic dimension increases, the Meissner stripe state is superseded as the ground state by states like those shown in panels (ii) and (iii). In these states, there are residual counter-flowing Meissner-like currents on the rows nearest the maxmimum-density rows; however, the dominant current instead now circulates around double-height plaquettes, with an alternating direction for neighbouring plaquettes. The appearance of these alternating plaquettes appears to be due to the influence of correlated pair tunneling interactions, while the residual Meissner-like currents likely stem from dipole-like terms [see Appendix B]. This indicates that the competition between different interaction effects sensitively depends on the synthetic dimension length for $\alpha=1/4$ with the parameters tested.  

Finally, for $\alpha \!=\!1/5$, we find ground states like those shown in Fig.~\ref{fig:hhmodelrwa1d5} for (i) $N_\lambda = 19$ and (ii) $N_{\lambda} = 34$ and other parameters as above. As expected, we observe that the density for these states is now modulated with a period of 5 sites along the synthetic dimension, again reflecting the larger magnetic unit cell. Continuing the above trends, we also observe for small system sizes [c.f. panel (i)] that the ground state is a type of Meissner stripe state. As compared to the similar states above, this Meissner stripe state now has two very low-density rows of sites carrying small opposed currents, reflecting the larger unit cell. We again find the same state when including only dipole-like interactions [see Appendix B]. 
However, for larger system sizes, we also again observe a more inhomogeneous ground state, as shown in panel (ii); here the density is strongly modulated also along the real dimension and the system is dominated by circulating currents (with alternating directions) rather than Meissner-like currents. Interestingly, we have also found that in this case, increasing the interaction strength, $g$, can instead lead to the re-emergence of the Meissner stripe phase (in this case however, most of the density vacates the center of the synthetic dimension), showing the rich competition between different effects.

 \section{Discussion and Conclusions} 
 \label{sec:conclusions}

In this paper, we have studied finite-size interacting hybrid Harper-Hofstadter systems in which one direction is provided by a real-space optical lattice, while the other is composed of a synthetic dimension of atomic trap states. For the quasi-1D HH ladder, we then found that including synthetic-dimension interactions did not significantly change the main features of the well-known Meissner and vortex-lattice phases, commonly also found in both single-particle and interacting systems. Instead the main effects of the unusual interactions were, firstly, to introduce an overall modulation of the particle density, causing particles to accumulate near the system edges, and, secondly, to de-favor the biased ladder phase which otherwise can emerge in interacting HH ladders. 

For the 2D HH model, we observed much more dramatic effects from the synthetic-dimension interactions, as these fundamentally changed the nature of the ground state. In particular, we identify a large class of states that we term ``Meissner stripe" states, which are characterized by Meissner-like counter-flowing currents along the real dimension and strong density modulations along the synthetic dimension, and which are present particularly for small synthetic-dimension lengths. In these cases, the structure of Meissner stripe state repeats along $\lambda$ and is controlled by the magnetic length, with the finite size of the system inducing a density envelope. This structure often leads these states to appear symmetric along $\lambda$, however, the presence of slight shifts of the center of mass in fact breaks such a symmetry. Interestingly, we also found that the ground state can change significantly as we change the length of the synthetic dimension, transitioning to states e.g. with some features more reminiscent of vortex-lattice-like states with  currents circulating around plaquettes, albeit in opposite and/or alternating directions. 
The interacting physics of these hybrid systems is therefore strongly system-size dependent; this is due to the inhomogeneity and long-range nature of the synthetic-dimension interaction terms as well as the competition between different subsets of these effects. 

 Although here we have focused on small finite systems, which will naturally be those most relevant to experiments, it will be interesting in the future to extend this work to larger synthetic dimensions, with the aim of beginning to explore the phase diagram in the thermodynamic limit. 

In the future, it will be important to go beyond this preliminary theoretical study to make a realistic experimental proposal, taking into account both experimental details and observables. For example, in Appendix D and E, we have studied the effects of  inhomogeneous hopping amplitudes, which arise from realistic shaking potentials used to couple the synthetic dimension, and have found e.g. similar Meissner stripe states to those shown above. Going further, it will be essential to also consider how these ground states could be prepared experimentally, and how robust they would be to the effects of heating from the Floquet driving used to create the synthetic dimension. While we have shown all ground states in this paper in the hybrid $x-\lambda$ plane,  experimental measurements will be made in the real $x-y$ plane. As discussed in Refs.~\cite{oliver2023bloch, reid}, it is possible to translate certain observables, such as the real-space center-of-mass position, into synthetic space, in order to infer information about the occupation of the atomic trap states from the real-space measurements. For states like the Meissner-stripe state discussed here, it is likely that the states will have very distinctive real-space density patterns as only a few trap states are significantly populated at a time [c.f. Sec.~\ref{sec:vortex_ground_states}]. 

Going further, another interesting avenue for future research is to go beyond mean-field physics to study the effects of strong interactions along the synthetic dimension, and the potential emergence of analogues of interesting strongly-correlated states, such as fractional quantum Hall states. For the HH ladder, the states we have found are very similar to those with other types of interactions, which may suggest that similar strongly-interacting physics could be found~\cite{piraud2015vortex,petrescu2015chiral, impertro2025strongly}. However, for the HH model, the states are fundamentally different from those previously studied, and can be sensitive to both system parameters and the strength of the interactions [see Appendix C]. These therefore may lead to novel types of strongly-correlated states in the presence of strong interactions. \\

 {\it Acknowledgements:} 
We thank Karel Van Acoleyen for helpful discussions. This work is supported by the Royal Society via grants UF160112, RGF\textbackslash{}EA\textbackslash{}180121 and RGF\textbackslash{}R1\textbackslash{}180071 and by the Engineering and Physical Sciences Research Council [grant numbers EP/W016141/1, EP/Y01510X/1, UKRI2226] and the CDT
in Topological Design [Grant No. EP/S02297X/1]. Work in
Brussels is supported by the ERC Grant LATIS, the EOS
project CHEQS, the Fonds de la Recherche Scientifique
(F.R.S.-FNRS) and the Fondation ULB. G.S. was funded by the Research Council of Finland through project n. 13354165 and the Italian Ministry of University and Research under the Rita Levi-Montalcini program. The computations described in this paper were performed using the University of Birmingham’s
Bluebear HPC service, which provides a High Performance
Computing service to the University’s research community.

\section*{Appendix A: Effective Floquet Hamiltonian}

\begin{figure}
\centering
\includegraphics[width=\columnwidth]{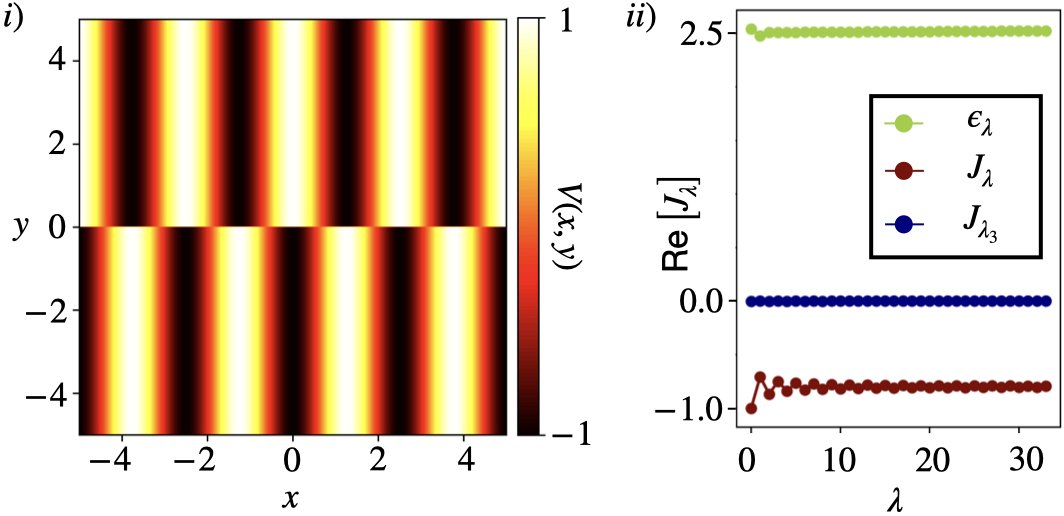}
\caption{\label{Fig:realistic_hopping} (i) The driving potential [Eq.~(\ref{Eq:realistic_potential})] at $t=0$ plotted for arbitrary parameters to show the structure of the modulation. The value of the magnetic flux $\varphi$ controls the length-scale of the modulation along the x direction. (ii) The nearest-neighbor hopping amplitudes found from the calculation of the Floquet Hamiltonian [Eq.~(\ref{Eq:logU})] for the driving potential from (i) with $x=0$, $\omega=\omega_d = 2\pi \times 166.5$Hz and $V_0 \!=\! -2.08$nK. We normalize these hopping amplitudes by the magnitude of the largest nearest neighbor hopping amplitude for comparison purposes.}
\end{figure}

As reviewed in the main text, the stroboscopic dynamics of a single particle in a shaken harmonic trap can be captured by an effective time-independent stroboscopic Floquet Hamiltonian [Eq.~(\ref{Eq:F})]~\cite{rudner2020floquetengineershandbook, PhysRevX.4.031027,Price2017}. To calculate this explicitly, we start from the 1D Hamiltonian 
\begin{equation}
\mathcal{H}_b= \frac{1}{2}\partial_y^2 + \frac{1}{2}y^2 + V(y, t),
\label{Eq:1dham}
\end{equation}
where $V(y, t)$ is the periodic driving potential with period $T$ and we have set $\hbar=m=\omega=1$~\cite{reid}. From this, we then numerically construct the stroboscopic evolution operator $\hat{U}(T, t_0=0)$ which evolves the wave function from $t=0$ up to $t=T$. The effective Hamiltonian, $\mathcal{H}_{\text{eff}}$, is then extracted as
\begin{equation}
    \mathcal{H}_{\text{eff}}\ \equiv \frac{i}{T}\log_e\left[ \hat{U}(T, t_0=0)\right].
    \label{Eq:logU}
\end{equation} 
In the main text, we focused on the driving potential [Eq.~(\ref{Eq:realistic_potential})] which is shown at $t=0$ in Fig.~\ref{Fig:realistic_hopping}(i). Experimentally, such a driving potential could be engineered by using, for example, a DMD \cite{Gauthier_2016, Amico_2021}. Using this potential to calculate Eq.~(\ref{Eq:logU}), we find that for any fixed value of $x$, the effective Floquet Hamiltonian can describe nearest-neighbor hopping processes, i.e. as denoted by $J_{\lambda}$: the matrix element between the $\lambda-1$ and $\lambda $ state. An additional complication is that an effective onsite potential arises from this driving, however, apart from the first few values of $\lambda$ these terms are essentially constant. All other terms of the Floquet Hamiltonian are negligible. These nearest-neighbor hopping amplitudes have the desired complex spatial dependence on $x$, i.e. $\exp[-i\varphi x]$. In Fig.~\ref{Fig:realistic_hopping}(ii), we plot these hopping amplitudes for $x\!=\!0$; as can be seen, these amplitudes oscillate around a constant value, with the magnitude of oscillations decaying quickly as $\lambda$ increases. Note that for illustrative purposes we have set $V_0 \!=\! -3.14$, such that these hopping amplitudes tend towards - 1 in the large $\lambda$ limit. In the main text, this behavior led us to approximate the nearest-neighbor hoppings as uniform, i.e. setting $J_{\lambda}\! =\! -t$. However, we have also explored the effects of keeping the full variation of these amplitudes, as discussed in Appendix D. 

\section*{Appendix B: Different interaction-term subsets}

To better understand the dominant interaction effects for the results shown in the main text, we have also numerically found the ground states for different subsets of synthetic-dimension interaction terms [c.f. Sec.~\ref{sec:inter}]. Here we include examples of these ground states, illustrating the main features referred to in the main text. 

\begin{figure}
    \centering
    \includegraphics[width=\linewidth]{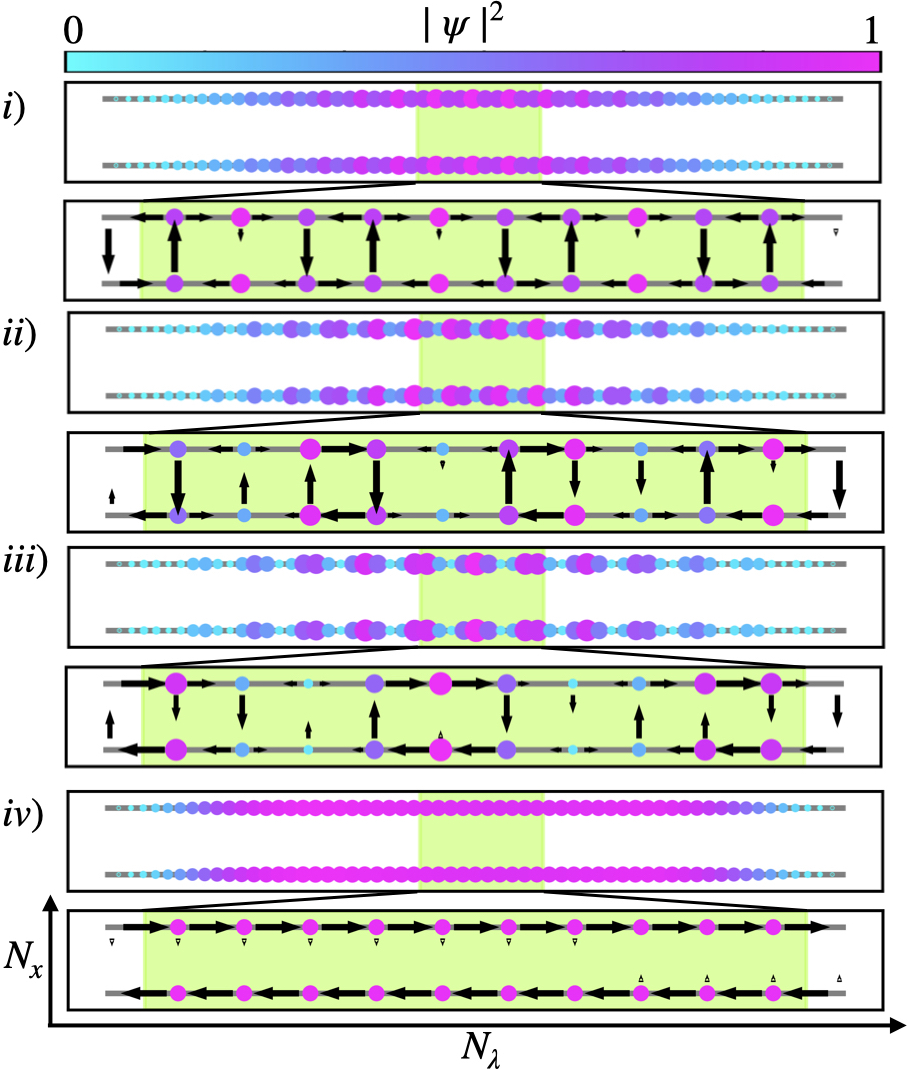}
    \caption{Example ground-states obtained for $\alpha=1/3$, $t=1$, $N_\lambda=59$ and only correlated pair tunneling synthetic-dimension interactions, with (i) $J_x=0.25$, (ii) $J_x=1$, (iii) $J_x=2$, (iv) $J_x=3.75$,
    i.e. parameters corresponding to the silver dots marked in Fig.~\ref{fig:phasediagram}(iv). As can be seen, panels (i)-(iii) are variants of vortex-lattice phases, without the biased ladder phase [c.f Fig.~\ref{fig:laddercontact}], while panel (iv) is a type of Meissner phase. }
    \label{fig:ladderelastic}
\end{figure}

\begin{figure}
    \centering
    \includegraphics[width=\linewidth]{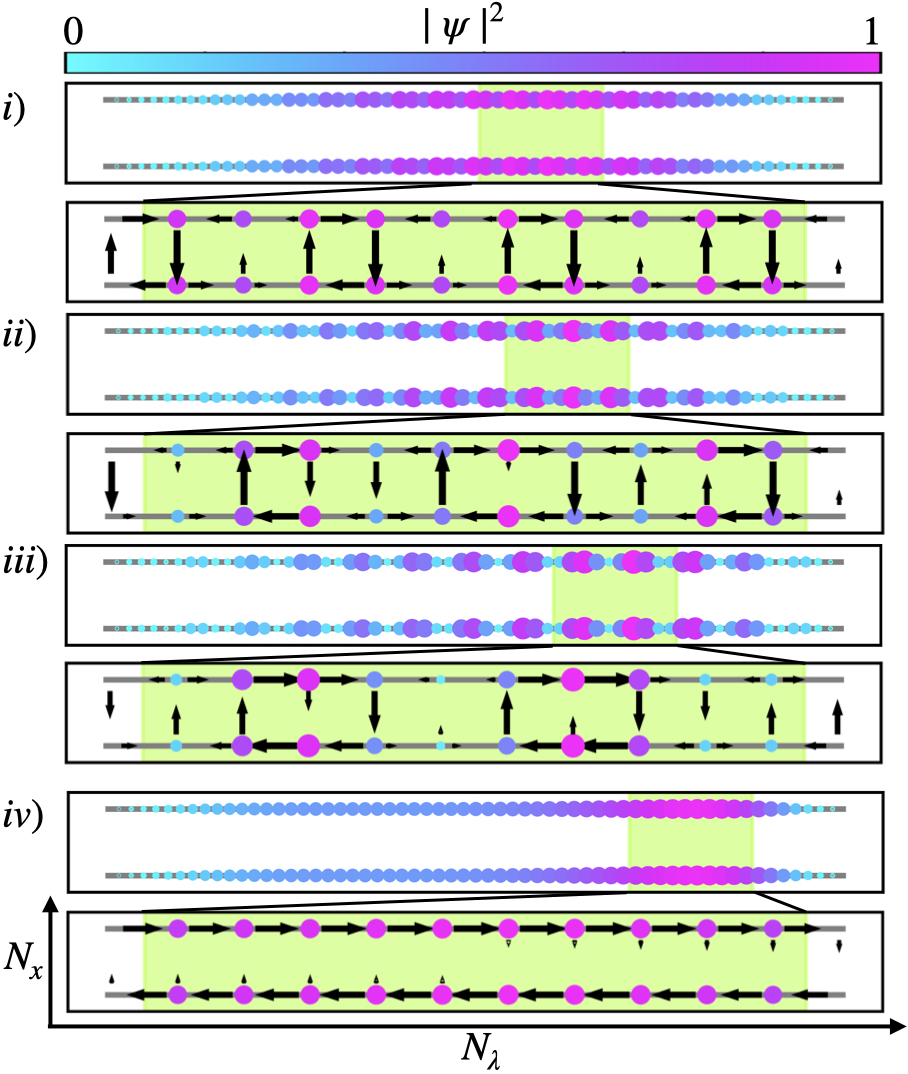}
    \caption{Example ground states obtained for $\alpha=1/3$, $t=1$, $N_\lambda=59$ and only dipole-like synthetic-dimension interactions, with (i) $J_x=0.25$, (ii) $J_x=1$, (iii) $J_x=2$, (iv) $J_x=3.75$,
    i.e. parameters corresponding to the silver dots marked in Fig.~\ref{fig:phasediagram}(v). These states have the same structure as those found in Fig.~\ref{fig:ladderelastic}, except now with the peak density shifted towards the upper edge of the synthetic dimension.}
    \label{fig:ladderdipole}
\end{figure}

Firstly, for the Harper-Hofstadter ladder [c.f. Sec.~\ref{sec:ladder}], examples of the ground states obtained with only correlated pair tunneling interactions are shown in Fig.~\ref{fig:ladderelastic}, for the same parameters as used in Sec.~\ref{sec:ladder}. As can be seen, these results are essentially the same as those found for standard contact interactions  [c.f. Fig.~\ref{fig:laddercontact}], except for the replacement of the biased ladder phase with the continuation of the vortex-lattice phase (e.g. as shown by $k=2$ in panel iii). For the same parameters, we also show the results with only dipole-like interactions in Fig.~\ref{fig:ladderdipole}. These states are qualitatively very similar to those found for correlated pair tunneling interactions above, except the maximum density is now clearly peaked towards the upper edge of the synthetic dimension. 

\begin{figure}
    \centering    \includegraphics[width=\linewidth]{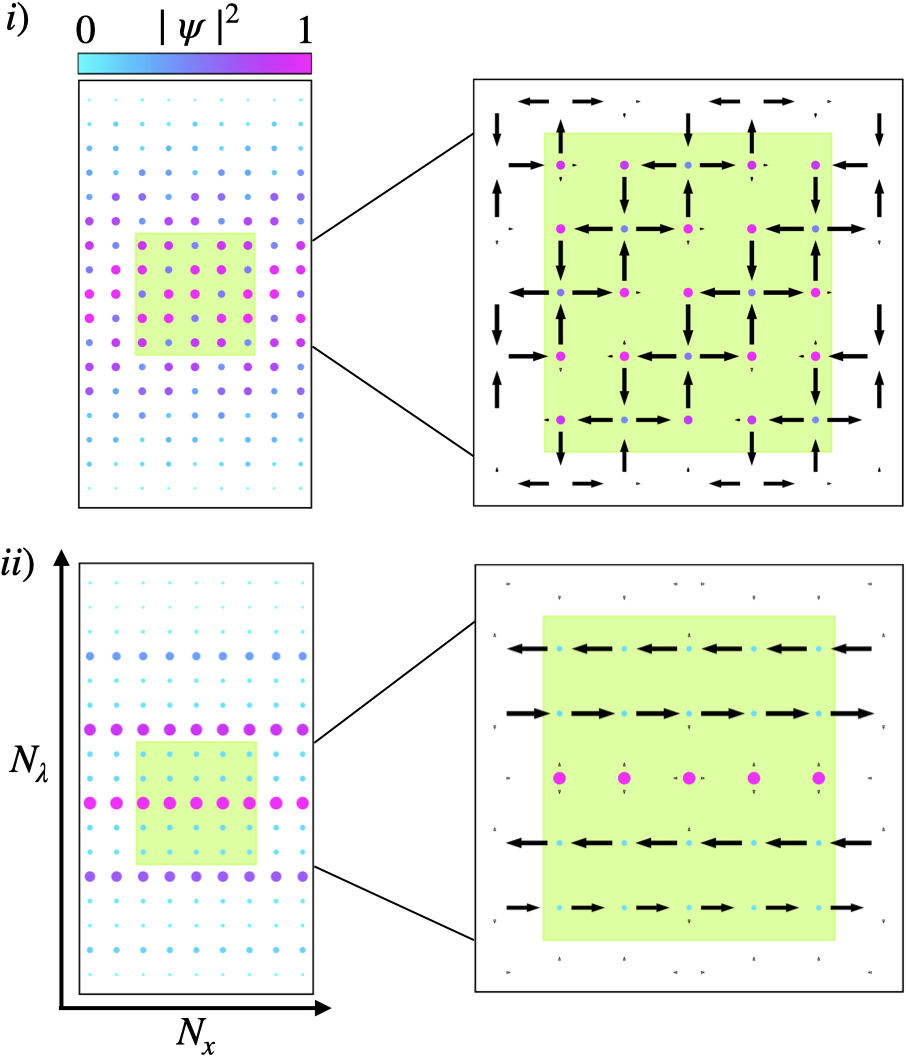}
    \caption{Ground states of the HH model with $\alpha\!=\!1/3$ with i) correlated pair tunneling and ii) dipole-like interactions with $N_\lambda \!=\! 17$ and other parameters as given in Fig.~\ref{fig:hhmodelrwa1d3}.}
    \label{fig:hhmodeldiffint1d3}
\end{figure}

Secondly, for the 2D Harper-Hofstadter model, we show example ground states with i) correlated pair tunneling and ii) dipole like interactions for $\alpha=1/3$ in Fig.~\ref{fig:hhmodeldiffint1d3}, for $N_\lambda=17$ and other parameters as in Fig.~\ref{fig:hhmodelrwa1d3}. As can be seen, correlated pair tunneling interactions lead to the vortex-lattice-like state previously observed for standard contact interactions [c.f. Fig.~\ref{fig:standardcontactinteractions}], while dipole-like interactions stabilize the Meissner stripe phase, seen for these parameters with the full RWA synthetic-dimension interactions [c.f. Fig.~\ref{fig:hhmodelrwa1d3}].

\begin{figure}
    \centering    \includegraphics[width=\linewidth]{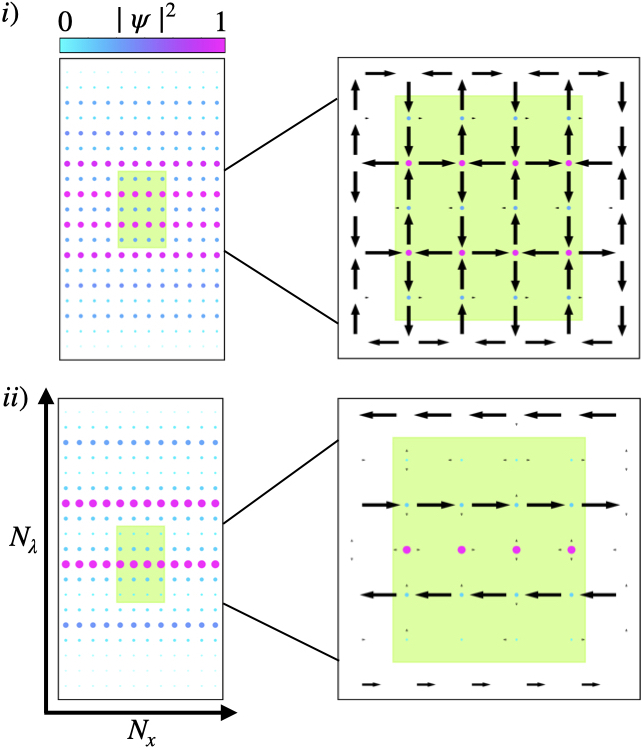}
    \caption{Ground states of the HH model with $\alpha\!=\!1/4$ with i) correlated pair tunneling and ii) dipole-like interactions with $N_\lambda \!=\! 19$ and other parameters as given in Fig.~\ref{fig:hhmodelrwa1d4}.}
    \label{fig:hhmodeldiffint1d4}
\end{figure}

\begin{figure}
    \centering    \includegraphics[width=\linewidth]{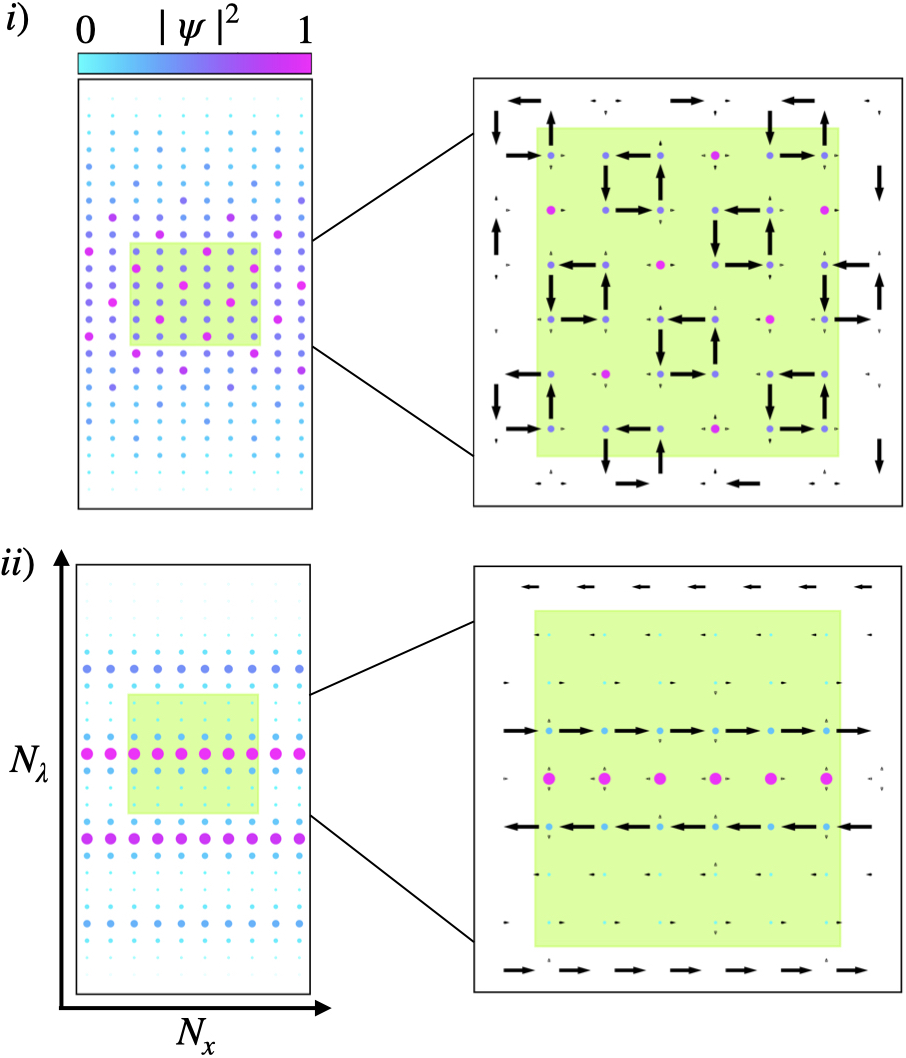}
    \caption{Ground states of the HH model with $\alpha\!=\!1/5$ with i) correlated pair tunneling and ii) dipole-like interactions with $N_\lambda \!=\! 24$ and other parameters as given in Fig.~\ref{fig:hhmodelrwa1d5}.}
    \label{fig:hhmodeldiffint1d5}
\end{figure}

Thirdly, the corresponding results for the HH model with $\alpha=1/4$ and i) correlated pair tunneling and ii) dipole-like interactions are shown in Fig.~\ref{fig:hhmodeldiffint1d4}, for $N_\lambda=19$ and other parameters as in Fig.~\ref{fig:hhmodelrwa1d4}. Again, we can see here that dipole-like interactions lead to  the Meissner stripe phase, which is also found for full RWA synthetic-dimension interactions with these parameters [c.f. Fig.~\ref{fig:hhmodelrwa1d4}]. However, here the correlated pair tunneling interactions do not lead to the standard contact interaction result [c.f. Fig.~\ref{fig:standardcontactinteractions}], but instead to alternating currents circulating around double-height plaquettes. These types of structures begin to emerge for RWA synthetic-dimension interactions in larger system sizes, suggesting that correlated pair tunneling interactions may become more important as the length of the synthetic dimension grows. 

Finally, example ground states for the HH model with $\alpha=1/5$ and i) correlated pair tunneling and ii) dipole-like interactions are shown in Fig.~\ref{fig:hhmodeldiffint1d5}, for $N_\lambda=24$ and other parameters as in Fig.~\ref{fig:hhmodelrwa1d5}. Here we again observe that the correlated pair tunneling interactions reproduce the standard contact interaction result  [c.f. Fig.~\ref{fig:standardcontactinteractions}], while the dipole-like interactions lead to the Meissner stripe phase [c.f. Fig.~\ref{fig:hhmodelrwa1d4}].

\section*{Appendix C: Changing the interaction strength}

We have repeated many of our numerical calculations for a range of interaction strengths, $g\in\left(0, 4\right]$. In most cases, we find states with the same types of current distributions as those shown in the main text, albeit with changes in the overall density profile. As expected, stronger interactions cause particles to spread out more, redistributing the particle density away from the central peak and towards the edges of the system. As a general trend, we also observe that the center-of-mass of the density moves upwards along the synthetic dimension; this reflects the inhomogeneity of the interactions along the synthetic dimension and is reminiscent of the density shifts observed in the HH ladder [c.f. Sec.~\ref{sec:ladder}]. Note that because of the underlying periodic density modulations along the synthetic dimension (from the magnetic unit cell), the position of the peak density does not shift smoothly but jumps discontinuously. Consequently, we also sometimes find several states that we cannot distinguish in energy within our numerical accuracy, with peak densities in different positions but otherwise all the same qualitative features.     

\begin{figure}
    \centering    \includegraphics[width=\linewidth]{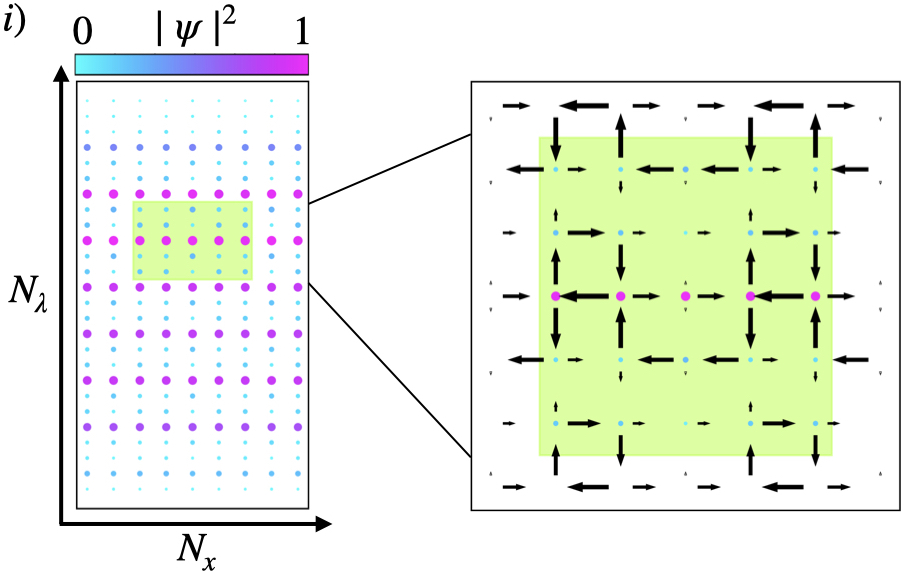}
    \caption{Ground state of the HH model with a stronger mean-field interaction strength, $g=4$, for full RWA synthetic-dimension interactions and $\alpha=1/3$, $N_\lambda=26$, $N_x=9$ and other parameters as in the main text. For these parameters, we find that increasing $g$ changes the ground state from a Meissner stripe phase [c.f. Fig.~\ref{fig:hhmodelrwa1d3}i)] to that shown here.}
    \label{fig:hhmodelg}
\end{figure}

However, for strong enough interactions, we also can see different types of states emerge for certain cases; for example, for $\alpha=1/3$, $N_\lambda=26$ and $g \gtrsim 3.6$, we find states like that shown in Fig.~\ref{fig:hhmodelg}, rather than the Meissner stripe phase found for smaller interaction strength with these parameters [c.f. Fig.~\ref{fig:hhmodelrwa1d3}i)]. This new state has the same structure of local currents as that previously found as a degeneracy for another system size [c.f. Fig.~\ref{fig:hhmodelrwa1d3}ii)], but with the density more spread out along the synthetic dimension due to the stronger interactions. This sensitivity to the interaction strength reflects the complexity of the interplay between different interaction terms in these small finite systems.   

\section*{Appendix D: Non-uniform Hopping Amplitudes}
\label{Appendix:Non_uniformhopping}
\begin{figure}
\centering
\includegraphics[width=\columnwidth]{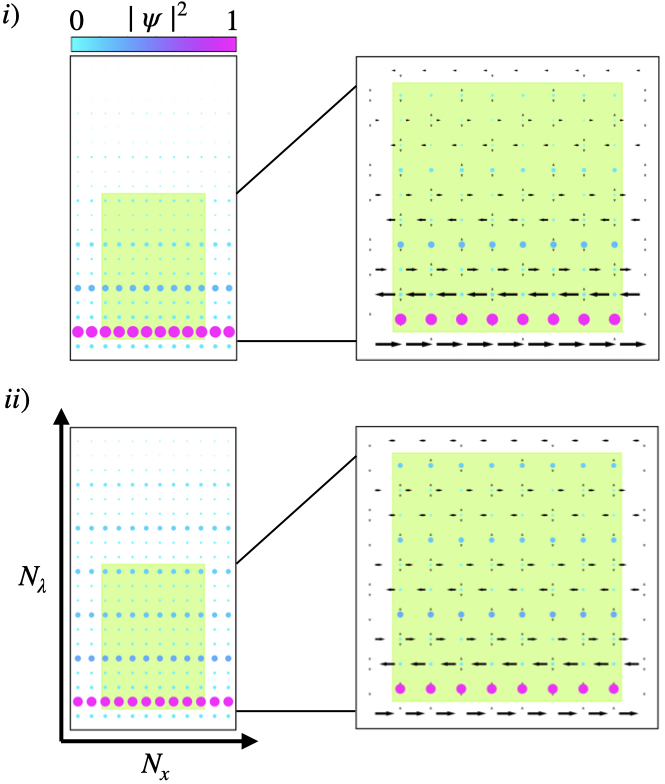}
\caption{\label{fig:gvsgroundstate} Ground states of the HH model with the inhomogeneous hopping amplitudes given in Fig.~\ref{Fig:realistic_hopping} with i) $g = 0.1$  and ii) $g=4.0$ for $\alpha =1/3$, $N_x = 12$ and $N_{\lambda}=20$, and all other parameters as in the main text.  These are Meissner stripe phases like those obtained previously for uniform hoppings [c.f. Fig.~\ref{fig:hhmodelrwa1d3} i], albeit with an overall downwards density shift due to the inhomogenity of these hopping amplitudes. }
\end{figure}

In the main text, we approximate the nearest-neighbor hopping amplitudes in the effective Floquet Hamiltonian to be uniform along the synthetic dimension for simplicity. For the specific driving potential given by Eq.~(\ref{Eq:realistic_potential}), this is well-justified for large $\lambda$ as shown in Appendix A. If we instead include the inhomogeneous hopping amplitudes for this case, then we obtain ground states such as those shown in Fig.~\ref{fig:gvsgroundstate} with  (i) $g\!=\! 0.1$ and (ii) $g\!=\! 4.0$ for $\alpha=1/3$, $N_x=12$ and $N_\lambda=20$ and all other parameters as given in the main text and in Fig.~\ref{Fig:realistic_hopping}. In this case all other hopping terms are negligible. Note that due to the non-uniformity of the hopping amplitudes, we have also had to modify the definition of the local current operator used to calculate the currents to:
\begin{equation}
J_{n\lambda}^{\uparrow} = -\left[iJ^{*}_{\lambda}e^{i\varphi n}\psi^{\dagger}_{n,\lambda}\psi_{n, \lambda-1}  + \text{h.c.}\right].
\end{equation}

As can be seen, in Fig.~\ref{fig:gvsgroundstate}, for these parameters, the ground states are Meissner stripe phases, similar to those found for uniform hoppings [c.f.~Fig.~\ref{fig:hhmodelrwa1d3}i)]. The main difference is that, in this case, the inhomogeneity of the hopping amplitudes causes the density distribution to shift down along the synthetic dimension by an amount which increases with the interaction strength.     

\begin{figure}
    \centering
    \includegraphics[width=\columnwidth]{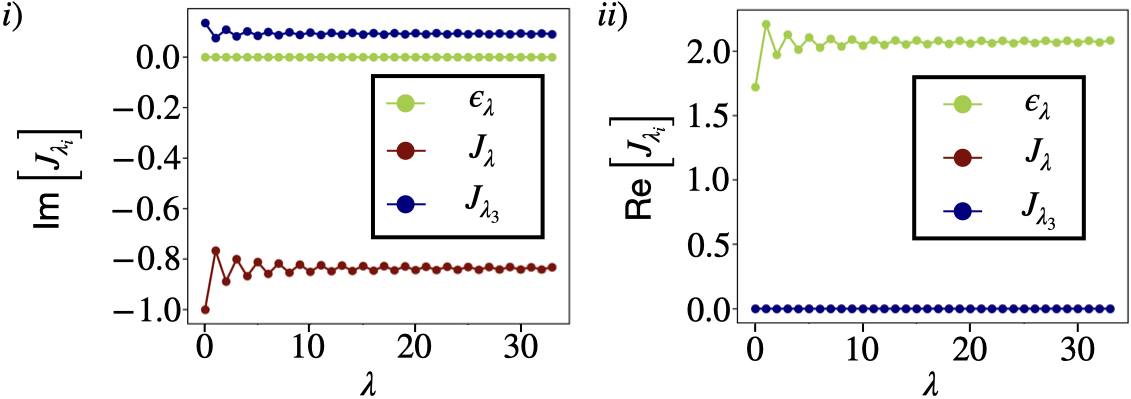}
    \caption{The i) imaginary and ii) real parts of the onsite (denoted by $\epsilon_\lambda$), nearest and next-next-nearest neighbor hopping amplitudes shown in green, red and blue respectively, for the effective Floquet Hamiltonian as calculated for the alternative driving potential [Eq.~(\ref{Eq:expdrive})] for $x\!=\!0$. Here we have chosen $ \omega_D \!=\!\omega\!=\! 2\pi \times 166.5$Hz and $V_0 = 3.9$nK to obtain the structure of the hopping amplitudes, however, we normalize all hopping amplitudes by the magnitude of the largest of the nearest neighbor hopping amplitudes so that we can compare with parameters utilized in the rest of the paper.}
    \label{fig:exppotentialgrounds}
\end{figure}

\section*{Appendix E: Alternative Driving Scheme}

\begin{figure}
\centering
\includegraphics[width=\columnwidth]{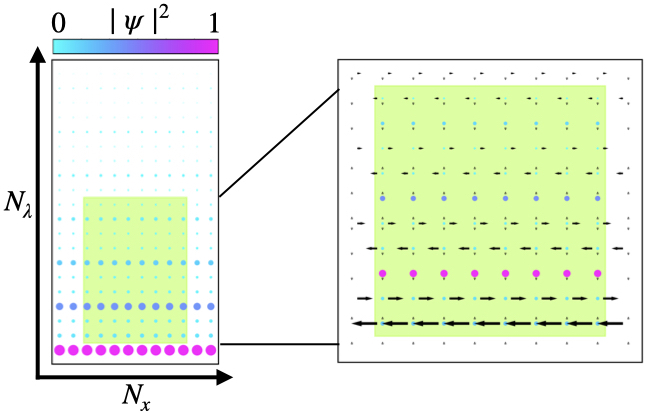}
\caption{\label{fig:Experimentresults} Ground states of the HH model with the hopping amplitudes taken from Fig.~\ref{fig:exppotentialgrounds} with $g=0.1$, $\alpha=1/3$, $N_x = 12$ and $N_{\lambda}=20$ and all other parameters as in the main text. These are again Meissner stripe states similar to those shown in Figs.~\ref{fig:gvsgroundstate} and~\ref{fig:hhmodelrwa1d3}i).}
\end{figure}

In the main text, we have focused on a particular type of time-periodic potential [Eq.~(\ref{Eq:realistic_potential})]; however, many other choices for the driving scheme can also efficiently couple the atomic trap states into a synthetic dimension. For example, in the recent experiment of Ref.~\cite{oliver2023bloch} the following type of driving potential was utilized,
\begin{equation}
V_{\text{exp}}(x, y, t) = V_0 \Theta(y\sin(\omega_D t +\varphi x)).
\label{Eq:expdrive}
\end{equation}
The effective Floquet Hamiltonian for this driving can again be calculated using Eq.~(\ref{Eq:logU}) using parameters described in Fig.~\ref{fig:exppotentialgrounds}. The resulting largest matrix elements are plotted in Fig.~\ref{fig:exppotentialgrounds}, corresponding to i) the imaginary and ii) the  real parts of the onsite ({\it green}), nearest-neighbor ({\it red}) and next-next-nearest-neighbor ({\it blue}) processes. As before the hopping amplitudes presented here are normalized by the magnitude of the largest nearest neighbor hopping term. Other hopping terms are much smaller and are henceforth neglected.

We can therefore construct an extended single-particle Hamiltonian given by
\begin{eqnarray}
{}&\mathcal{H}_{\text{exp}} =  \sum_{n,\lambda}  -J_x \psi_{n-1, \lambda}^{\dagger}\psi_{n,\lambda} + J_{\lambda}e^{-i \varphi n}\psi_{n, \lambda-1}^{\dagger}\psi_{n, \lambda} + \nonumber\\ &J_{\lambda_3}e^{-i 3\varphi n}\psi_{n, \lambda-3}^{\dagger}\psi_{n,\lambda} + \text{h.c.} , \label{Hexp}
\end{eqnarray}
where $J_{\lambda_3}$ now corresponds to the next-next-nearest neighbor hopping amplitudes. Note that these processes are accompanied by a hopping phase which is three times larger than that of the nearest-neighbor processes. We have checked this dependence numerically by substituting different values of $x$ in Eq.~(\ref{Eq:expdrive}) when calculating the 1D Floquet Hamiltonian. The local density evolves as
\begin{equation}
\dot{\mathcal{P}}_{n, \lambda} = i \left[ \mathcal{H}_{\text{exp}}, \psi_{n,\lambda}^{\dagger}\psi_{n,\lambda} \right],
\end{equation} 
which leads to a modified current flowing along the synthetic dimension as
\begin{eqnarray}
    \mathcal{J}_{n, \lambda}^{\uparrow} = {}&-[iJ_{\lambda}^{*}e^{i \varphi n}\psi_{n, \lambda}^{\dagger}\psi_{n,\lambda-1} + \\ & iJ^{*}_{\lambda_3}e^{i3\varphi n}\psi_{n, \lambda}^{\dagger}\psi_{n, \lambda-3} + \text{h.c.}].
    \label{Eq:modcurrents}
\end{eqnarray} 
Note that as mentioned in the main text, we have neglected the current contributions stemming from the interaction terms in the full Hamiltonian as these are much smaller in magnitude. We then numerically find the ground state for the combined Hamiltonian given by $\mathcal{H}_{\text{exp}}+ \mathcal{H}_{\text{int}}$, where the latter term corresponds to the full RWA synthetic-dimension interactions [c.f. Sec.~\ref{sec:inter}]. Examples of the resulting density and current distributions are shown in Fig.~\ref{fig:Experimentresults} for (i)$g\!=\! 0.1$ and (ii) $g\!=\! 4.0$ for $\alpha=1/3$, $N_x=12$ and $N_\lambda=29$ and all other parameters as given in the main text and in Fig.~\ref{fig:exppotentialgrounds}. As can be seen, for these parameters, the ground states are again Meissner stripe phases, similar to those found for both uniform hoppings and non-uniform hoppings [c.f.~Fig.~\ref{fig:hhmodelrwa1d3}i) and Fig.~\ref{fig:gvsgroundstate} respectively].

\appendix
\bibliography{bibliography}

@article{PhysRevA.83.013612,
  title = {Bogoliubov theory of interacting bosons on a lattice in a synthetic magnetic field},
  author = {Powell, Stephen and Barnett, Ryan and Sensarma, Rajdeep and Das Sarma, Sankar},
  journal = {Phys. Rev. A},
  volume = {83},
  issue = {1},
  pages = {013612},
  numpages = {19},
  year = {2011},
  month = {Jan},
  publisher = {American Physical Society},
  doi = {10.1103/PhysRevA.83.013612},
  url = {https://link.aps.org/doi/10.1103/PhysRevA.83.013612}
}

@article{oliver2023bloch,
  title={Bloch oscillations along a synthetic dimension of atomic trap states},
  author={Oliver, Christopher and Smith, Aaron and Easton, Thomas and Salerno, Grazia and Guarrera, Vera and Goldman, Nathan and Barontini, Giovanni and Price, Hannah M},
  journal={Physical Review Research},
  volume={5},
  number={3},
  pages={033001},
  year={2023},
  publisher={APS}
}

@article{Price2017,
  title = {Synthetic dimensions for cold atoms from shaking a harmonic trap},
  author = {Price, H.M. and Ozawa, T. and Goldman, N.},
  journal = {Phys. Rev. A},
  volume = {95},
  issue = {2},
  pages = {023607},
  numpages = {17},
  year = {2017},
  month = {Feb},
  publisher = {American Physical Society},
  url = {https://link.aps.org/doi/10.1103/PhysRevA.95.023607}
}

@article{PhysRevX.4.031027,
  title = {Periodically Driven Quantum Systems: Effective Hamiltonians and Engineered Gauge Fields},
  author = {Goldman, N. and Dalibard, J.},
  journal = {Phys. Rev. X},
  volume = {4},
  issue = {3},
  pages = {031027},
  numpages = {29},
  year = {2014},
  month = {8},
  publisher = {American Physical Society},
  doi ={10.1103/PhysRevX.4.031027},
  url = {https://link.aps.org/doi/10.1103/PhysRevX.4.031027}
}

@article{Boada2012,
  title = {Quantum Simulation of an Extra Dimension},
  author = {Boada, O. and Celi, A. and Latorre, J. I. and Lewenstein, M.},
  journal = {Phys. Rev. Lett.},
  volume = {108},
  issue = {13},
  pages = {133001},
  numpages = {6},
  year = {2012},
  month = {3},
  publisher = {American Physical Society},
}

@article{Celi2014,
  title = {Synthetic Gauge Fields in Synthetic Dimensions},
  author = {Celi, A. and Massignan, P. and Ruseckas, J. and Goldman, N. and Spielman, I. B. and Juzeli\ifmmode \bar{u}\else \={u}\fi{}nas, G. and Lewenstein, M.},
  journal = {Phys. Rev. Lett.},
  volume = {112},
  issue = {4},
  pages = {043001},
  numpages = {6},
  year = {2014},
  month = {Jan},
  publisher = {American Physical Society},
  url = {https://link.aps.org/doi/10.1103/PhysRevLett.112.043001}
}

@article{Cooper2019,
  title = {Topological bands for ultracold atoms},
  author = {Cooper, N. R. and Dalibard, J. and Spielman, I. B.},
  journal = {Rev. Mod. Phys.},
  volume = {91},
  issue = {1},
  pages = {015005},
  numpages = {55},
  year = {2019},
  month = {Mar},
  publisher = {American Physical Society},
  url = {https://link.aps.org/doi/10.1103/RevModPhys.91.015005}
}

@article{Sundar2018,
  title = {Synthetic dimensions in ultracold polar molecules.},
  author = {Sundar, B. and Gadway, B. and Hazzard, K.R.A. },
  journal = {Sci. Rep.},
  volume = {8},
  pages = {3422},
  year = {2018},
  url = {https://www.nature.com/articles/s41598-018-21699-x}
}

@article{Sundar2019,
  title = {Strings of ultracold molecules in a synthetic dimension},
  author = {Sundar, B. and Thibodeau, M. and Wang, Z. and Gadway, B. and Hazzard, K.R. A.},
  journal = {Phys. Rev. A},
  volume = {99},
  issue = {1},
  pages = {013624},
  numpages = {11},
  year = {2019},
  month = {Jan},
  publisher = {American Physical Society},
  url = {https://link.aps.org/doi/10.1103/PhysRevA.99.013624}
}

@article{Wang2021,
  title = {Generating arbitrary topological windings of a non-Hermitian band},
  author = {Wang, K. and Dutt, A. and Yang, Y.Y. and Wojcik, C.C. and Vu\v{c}kovi\'{c}, J. and Fan, S.},
  journal = {Science},
  volume = {371},
  pages = {6535},
  year = {2021},
  url = {https://www.science.org/doi/full/10.1126/science.abf6568}
}

@article{lustig2019photonic,
  title={Photonic topological insulator in synthetic dimensions},
  author={Lustig, E. and Weimann, S. and Plotnik, Y. and Lumer, Y. and Bandres, M.A. and Szameit, A. and Segev, M.},
  journal={Nature},
  volume={567},
  number={7748},
  pages={356--360},
  year={2019},
  publisher={Nature Publishing Group},
  url = {https://www.nature.com/articles/s41586-019-0943-7/}
}

@article{lustig2021topological,
  title={Topological photonics in synthetic dimensions},
  author={Lustig, Eran and Segev, Mordechai},
  journal={{Advances} in Optics and Photonics},
  volume={13},
  number={2},
  pages={426--461},
  year={2021},
  publisher={Optical Society of America}
}

@article{dutt2020single,
  title={A single photonic cavity with two independent physical synthetic dimensions},
  author={Dutt, A. and Lin, Q. and Yuan, L. and Minkov, M. and Xiao, M. and Fan, S.},
  journal={Science},
  volume={367},
  number={6473},
  pages={59--64},
  year={2020},
  publisher={American Association for the Advancement of Science},
  url = {https://www.science.org/doi/10.1126/science.aaz3071}
}

@article{Mancini2015,
author = {M. Mancini  and G. Pagano  and G. Cappellini  and L. Livi  and M. Rider  and J. Catani  and C. Sias  and P. Zoller  and M. Inguscio  and M. Dalmonte  and L. Fallani },
title = {Observation of chiral edge states with neutral fermions in synthetic {H}all ribbons},
journal = {Science},
volume = {349},
number = {6255},
pages = {1510-1513},
year = {2015},
URL = {https://www.science.org/doi/abs/10.1126/science.aaa8736},
}

@article{Stuhl2015,
author = {B. K. Stuhl  and H.-I. Lu  and L. M. Aycock  and D. Genkina  and I. B. Spielman },
title = {Visualizing edge states with an atomic {B}ose gas in the quantum {H}all regime},
journal = {Science},
volume = {349},
number = {6255},
pages = {1514-1518},
year = {2015},
URL = {https://www.science.org/doi/abs/10.1126/science.aaa8515},
}

@article{Livi2016,
  title = {Synthetic Dimensions and Spin-Orbit Coupling with an Optical Clock Transition},
  author = {Livi, L. F. and Cappellini, G. and Diem, M. and Franchi, L. and Clivati, C. and Frittelli, M. and Levi, F. and Calonico, D. and Catani, J. and Inguscio, M. and Fallani, L.},
  journal = {Phys. Rev. Lett.},
  volume = {117},
  issue = {22},
  pages = {220401},
  numpages = {5},
  year = {2016},
  month = {Nov},
  publisher = {American Physical Society},
  url = {https://link.aps.org/doi/10.1103/PhysRevLett.117.220401}
}

@article{Kolkowitz2017,
  title = {Spin–orbit-coupled fermions in an optical lattice clock},
  author = {Kolkowitz, S. and Bromley, S. and Bothwell, T. and M. L. Wall and G. E. Marti and A. P. Koller and X. Zhang and A. M. Rey and J. Ye},
  journal = {Nature},
  volume = {542},
  pages = {66 - 70},
  year = {2017},
  url = {https://www.nature.com/articles/nature20811}
}

@article{Chalopin2020,
  title = {Probing chiral edge dynamics and bulk topology of a synthetic Hall system.},
  author = {Chalopin, T. and Satoor, T. and Evrard, A. and Mahkalov, V. and Dalibard, J. and Lopes, R. and Nascimbene, S.},
  journal = {Nat. Phys.},
  volume = {16},
  pages = {1017–1021},
  year = {2020},
  url = {https://www.nature.com/articles/s41567-020-0942-5}
}

@article{Kanungo2021,
  title={Realizing topological edge states with Rydberg-atom synthetic dimensions},
  author={Kanungo, SK and Whalen, JD and Lu, Y and Yuan, M and Dasgupta, S and Dunning, FB and Hazzard, KRA and Killian, TC},
  journal={Nat. {Commun.}},
  volume={13},
  number={1},
  pages={972},
  year={2022},
  publisher={Nature Publishing Group UK London}
}

@article{lienhard2020realization,
  title={Realization of a density-dependent {P}eierls phase in a synthetic, spin-orbit coupled {R}ydberg system},
  author={Lienhard, V. and Scholl, P. and Weber, S. and Barredo, D. and de L{\'e}s{\'e}leuc, S. and Bai, R. and Lang, N. and Fleischhauer, M. and B{\"u}chler, H.P. and Lahaye, T. and Browaeys, A.},
  journal={Phys. Rev. X},
  volume={10},
  number={2},
  pages={021031},
  year={2020},
  publisher={APS},
  url = {https://journals.aps.org/prx/abstract/10.1103/PhysRevX.10.021031}
}

@article{chen2021real,
  title={Real-time observation of frequency {B}loch oscillations with fibre loop modulation},
  author={Chen, H. and Yang, N. and Qin, C. and Li, W. and Wang, B. and Han, T. and Zhang, C. and Liu, W. and Wang, K. and Long, H. and Zhang, X. and Peixiang, Lu.},
  journal={Light Sci. Appl.},
  volume={10},
  number={1},
  pages={1--9},
  year={2021},
  publisher={Nature Publishing Group},
  url = {https://www.nature.com/articles/s41377-021-00494-w}
}

@article{meier2018observation,
  title={Observation of the topological {A}nderson insulator in disordered atomic wires},
  author={Meier, E.J. and An, F.A. and Dauphin, A. and Maffei, M. and Massignan, P. and Hughes, T.L. and Gadway, B.},
  journal={Science},
  volume={362},
  number={6417},
  pages={929--933},
  year={2018},
  publisher={American Association for the Advancement of Science},
  url = {https://www.science.org/doi/10.1126/science.aat3406}
}

@article{Gadway2016,
  title = {Observation of the topological soliton state in the {S}u–{S}chrieffer–{H}eeger model},
  author = {Meier, E. and An, F. and Gadway, B.},
  journal = {Nat. Commun.},
  volume = {7},
  pages = {13986},
  year = {2016},
  url = {https://www.nature.com/articles/ncomms13986}
}

@article{Viebahn2019,
  title = {Matter-Wave Diffraction from a Quasicrystalline Optical Lattice},
  author = {Viebahn, K. and Sbroscia, M. and Carter, E. and Yu, J. and Schneider, U.},
  journal = {Phys. Rev. Lett.},
  volume = {122},
  issue = {11},
  pages = {110404},
  numpages = {6},
  year = {2019},
  month = {Mar},
  publisher = {American Physical Society},
  url = {https://link.aps.org/doi/10.1103/PhysRevLett.122.110404}
}

@article{cai2019experimental,
  title={Experimental observation of momentum-space chiral edge currents in room-temperature atoms},
  author={Cai, H. and Liu, J. and Wu, J. and He, Y. and Zhu, S. and Zhang, J. and Wang, D.},
  journal={Phys. Rev. Lett.},
  volume={122},
  number={2},
  pages={023601},
  year={2019},
  publisher={APS},
  url = {https://journals.aps.org/prl/abstract/10.1103/PhysRevLett.122.023601}
}

@article{barbiero2020bose,
  title={Bose-Hubbard physics in synthetic dimensions from interaction Trotterization},
  author={Barbiero, Luca and Chomaz, Lauriane and Nascimbene, S and Goldman, N},
  journal={Phys. Rev. Research},
  volume={2},
  number={4},
  pages={043340},
  year={2020},
  publisher={APS}
}

@article{balvcytis2021synthetic,
  title={Synthetic dimension band structures on a Si CMOS photonic platform},
  author={Bal{\v{c}}ytis, Armandas and Ozawa, Tomoki and Ota, Yasutomo and Iwamoto, Satoshi and Maeda, Jun and Baba, Toshihiko},
  journal={Science {Advances}},
  volume={8},
  number={4},
  pages={eabk0468},
  year={2022},
  publisher={American Association for the Advancement of Science}
}

@article{kang2020creutz,
  title={Creutz ladder in a resonantly shaken 1D optical lattice},
  author={Kang, Jin Hyoun and Han, Jeong Ho and Shin, Yi},
  journal={New Journal of Physics},
  volume={22},
  number={1},
  pages={013023},
  year={2020},
  publisher={IOP Publishing},
  url = {https://iopscience.iop.org/article/10.1088/1367-2630/ab61d7}
}

@article{Salerno2019,
  title = {Quantized {H}all Conductance of a Single Atomic Wire: A Proposal Based on Synthetic Dimensions},
  author = {Salerno, G. and Price, H. M. and Lebrat, M. and H\"ausler, S. and Esslinger, T. and Corman, L. and Brantut, J.-P. and Goldman, N.},
  journal = {Phys. Rev. X},
  volume = {9},
  issue = {4},
  pages = {041001},
  numpages = {21},
  year = {2019},
  month = {Oct},
  publisher = {American Physical Society},
  url = {https://link.aps.org/doi/10.1103/PhysRevX.9.041001}
}

@article{price2015,
  title = {Four-Dimensional Quantum {H}all Effect with Ultracold Atoms},
  author = {Price, H. M. and Zilberberg, O. and Ozawa, T. and Carusotto, I. and Goldman, N.},
  journal = {Phys. Rev. Lett.},
  volume = {115},
  issue = {19},
  pages = {195303},
  numpages = {6},
  year = {2015},
  month = {Nov},
  publisher = {American Physical Society},
  url = {https://link.aps.org/doi/10.1103/PhysRevLett.115.195303}
}

@article{bouhiron2024realization,
  title={Realization of an atomic quantum Hall system in four dimensions},
  author={Bouhiron, Jean-Baptiste and Fabre, Aur{\'e}lien and Liu, Qi and Redon, Quentin and Mittal, Nehal and Satoor, Tanish and Lopes, Raphael and Nascimbene, Sylvain},
  journal={Science},
  volume={384},
  number={6692},
  pages={223--227},
  year={2024},
  publisher={American Association for the Advancement of Science}
}

@article{ozawa2017synthetic,
  title={Synthetic dimensions with magnetic fields and local interactions in photonic lattices},
  author={Ozawa, Tomoki and Carusotto, Iacopo},
  journal={Phys. Rev. Lett.},
  volume={118},
  number={1},
  pages={013601},
  year={2017}
}

@article{An2021,
  title = {Nonlinear Dynamics in a Synthetic Momentum-State Lattice},
  author = {An, Fangzhao Alex and Sundar, Bhuvanesh and Hou, Junpeng and Luo, Xi-Wang and Meier, Eric J. and Zhang, Chuanwei and Hazzard, Kaden R. A. and Gadway, Bryce},
  journal = {Phys. Rev. Lett.},
  volume = {127},
  issue = {13},
  pages = {130401},
  numpages = {7},
  year = {2021},
  month = {Sep},
  publisher = {American Physical Society},
  doi = {10.1103/PhysRevLett.127.130401},
  url = {https://link.aps.org/doi/10.1103/PhysRevLett.127.130401}
}

@article{chen2024interaction,
  title={Interaction-driven breakdown of Aharonov--Bohm caging in flat-band Rydberg lattices},
  author={Chen, Tao and Huang, Chenxi and Velkovsky, Ivan and Ozawa, Tomoki and Price, Hannah and Covey, Jacob P and Gadway, Bryce},
  journal={arXiv:2404.00737},
year=2024}

@article{chen2024strongly,
  title={Strongly interacting Rydberg atoms in synthetic dimensions with a magnetic flux},
  author={Chen, Tao and Huang, Chenxi and Velkovsky, Ivan and Hazzard, Kaden RA and Covey, Jacob P and Gadway, Bryce},
  journal={Nat. {Commun.}},
  volume={15},
  number={1},
  pages={2675},
  year={2024},
  publisher={Nature Publishing Group UK London}
}

@article{PRXQuantum.5.010310,
  title = {Two-Dimensional Momentum State Lattices},
  author = {Agrawal, Shraddha and Paladugu, Sai Naga Manoj and Gadway, Bryce},
  journal = {PRX Quantum},
  volume = {5},
  issue = {1},
  pages = {010310},
  numpages = {14},
  year = {2024},
  month = {Jan},
  publisher = {American Physical Society},
  doi = {10.1103/PRXQuantum.5.010310},
  url = {https://link.aps.org/doi/10.1103/PRXQuantum.5.010310}
}

@article{dinh2024reconfigurable,
  title={Reconfigurable synthetic dimension frequency lattices in an integrated lithium niobate ring cavity},
  author={Dinh, Hiep X and Bal{\v{c}}ytis, Armandas and Ozawa, Tomoki and Ota, Yasutomo and Ren, Guanghui and Baba, Toshihiko and Iwamoto, Satoshi and Mitchell, Arnan and Nguyen, Thach G},
  journal={{Commun.} Physics},
  volume={7},
  number={1},
  pages={185},
  year={2024},
  publisher={Nature Publishing Group UK London}
}

@article{leefmans2022topological,
  title={Topological dissipation in a time-multiplexed photonic resonator network},
  author={Leefmans, Christian and Dutt, Avik and Williams, James and Yuan, Luqi and Parto, Midya and Nori, Franco and Fan, Shanhui and Marandi, Alireza},
  journal={Nat. {Phys.}},
  volume={18},
  number={4},
  pages={442--449},
  year={2022},
  publisher={Nature Publishing Group UK London}
}

@article{balvcytis2022synthetic,
  title={Synthetic dimension band structures on a Si CMOS photonic platform},
  author={Bal{\v{c}}ytis, Armandas and Ozawa, Tomoki and Ota, Yasutomo and Iwamoto, Satoshi and Maeda, Jun and Baba, Toshihiko},
  journal={Science {Advances}},
  volume={8},
  number={4},
  pages={eabk0468},
  year={2022},
  publisher={American Association for the Advancement of Science}
}

@article{ehrhardt2023perspective,
  title={A perspective on synthetic dimensions in photonics},
  author={Ehrhardt, Max and Weidemann, Sebastian and Maczewsky, Lukas J and Heinrich, Matthias and Szameit, Alexander},
  journal={Laser \& Photonics Reviews},
  volume={17},
  number={7},
  pages={2200518},
  year={2023},
  publisher={Wiley Online Library}
}

@article{PhysRevLett.132.130601,
  title = {Realizing Synthetic Dimensions and Artificial Magnetic Flux in a Trapped-Ion Quantum Simulator},
  author = {Wang, Y. and Wu, Y.-K. and Jiang, Y. and Cai, M.-L. and Li, B.-W. and Mei, Q.-X. and Qi, B.-X. and Zhou, Z.-C. and Duan, L.-M.},
  journal = {Phys. Rev. Lett.},
  volume = {132},
  issue = {13},
  pages = {130601},
  numpages = {6},
  year = {2024},
  month = {Mar},
  publisher = {American Physical Society},
  doi = {10.1103/PhysRevLett.132.130601},
  url = {https://link.aps.org/doi/10.1103/PhysRevLett.132.130601}
}

@article{Roell2023,
  title = {Chiral edge dynamics and quantum Hall physics in synthetic dimensions with an atomic erbium Bose-Einstein condensate},
  author = {Roell, Roberto Vittorio and Laskar, Arif Warsi and Huybrechts, Franz Richard and Weitz, Martin},
  journal = {Phys. Rev. A},
  volume = {107},
  issue = {4},
  pages = {043302},
  numpages = {6},
  year = {2023},
  month = {Apr},
  publisher = {American Physical Society},
}

@article{lu2024probing,
  title={Probing the topological phase transition in the Su-Schrieffer-Heeger model using Rydberg-atom synthetic dimensions},
  author={Lu, Yi and Wang, Chuanyu and Kanungo, Soumya K and Dunning, F Barry and Killian, Thomas C},
  journal={arXiv:2404.18420},
  year ={2024},
}

@article{Dalibard_2011,
   title={Colloquium: Artificial gauge potentials for neutral atoms},
   volume={83},
   ISSN={1539-0756},
   url={http://dx.doi.org/10.1103/RevModPhys.83.1523},
   DOI={10.1103/revmodphys.83.1523},
   number={4},
   journal={Reviews of Modern Physics},
   publisher={American Physical Society (APS)},
   author={Dalibard, Jean and Gerbier, Fabrice and Juzeliūnas, Gediminas and Öhberg, Patrik},
   year={2011},
   month=nov, pages={1523–1543} }

@article{Galitski_2019,
   title={Artificial gauge fields with ultracold atoms},
   volume={72},
   ISSN={1945-0699},
   url={http://dx.doi.org/10.1063/PT.3.4111},
   DOI={10.1063/pt.3.4111},
   number={1},
   journal={Physics Today},
   publisher={AIP Publishing},
   author={Galitski, Victor and Juzeliūnas, Gediminas and Spielman, Ian B.},
   year={2019},
   month=jan, pages={38–44} }

@article{oliver2023artificial,
  title={Artificial gauge fields in the t-z mapping for optical pulses: Spatiotemporal wave packet control and quantum Hall physics},
  author={Oliver, Christopher and Mukherjee, Sebabrata and Rechstman, Mikael C and Carusotto, Iacopo and Price, Hannah M},
  journal={Science {Advances}},
  volume={9},
  number={42},
  pages={eadj0360},
  year={2023},
  publisher={American Association for the Advancement of Science}
}

@article{ozawa2019topological,
  title={Topological quantum matter in synthetic dimensions},
  author={Ozawa, Tomoki and Price, Hannah M},
  journal={Nature Reviews Physics},
  volume={1},
  number={5},
  pages={349--357},
  year={2019},
  publisher={Nature Publishing Group UK London}
}

@article{leonard2023realization,
  title={Realization of a fractional quantum Hall state with ultracold atoms},
  author={L{\'e}onard, Julian and Kim, Sooshin and Kwan, Joyce and Segura, Perrin and Grusdt, Fabian and Repellin, C{\'e}cile and Goldman, Nathan and Greiner, Markus},
  journal={Nature},
  volume={619},
  number={7970},
  pages={495--499},
  year={2023},
  publisher={Nature Publishing Group UK London}
}

@article{an2018correlated,
  title={Correlated dynamics in a synthetic lattice of momentum states},
  author={An, Fangzhao Alex and Meier, Eric J and Ang’ong’a, Jackson and Gadway, Bryce},
  journal={Physical review letters},
  volume={120},
  number={4},
  pages={040407},
  year={2018},
  publisher={APS}
}

@article{reid,
  title = {Topological chiral edge states in a synthetic dimension of atomic trap states},
  author = {Reid, David G. and Oliver, Christopher and Regan, Patrick and Smith, Aaron and Easton, Thomas and Salerno, Grazia and Barontini, Giovanni and Goldman, Nathan and Price, Hannah M.},
  journal = {Phys. Rev. A},
  volume = {111},
  issue = {3},
  pages = {033301},
  numpages = {20},
  year = {2025},
  month = {Mar},
  publisher = {American Physical Society},
  doi = {10.1103/PhysRevA.111.033301},
  url = {https://link.aps.org/doi/10.1103/PhysRevA.111.033301}
}

@article{chang2005coherent,
  title={Coherent spinor dynamics in a spin-1 Bose condensate},
  author={Chang, Ming-Shien and Qin, Qishu and Zhang, Wenxian and You, Li and Chapman, Michael S},
  journal={Nature physics},
  volume={1},
  number={2},
  pages={111--116},
  year={2005},
  publisher={Nature Publishing Group UK London}
}

@article{zhou2023observation,
  title={Observation of universal Hall response in strongly interacting Fermions},
  author={Zhou, T-W and Cappellini, G and Tusi, D and Franchi, L and Parravicini, J and Repellin, C and Greschner, S and Inguscio, Massimo and Giamarchi, T and Filippone, M and others},
  journal={Science},
  volume={381},
  number={6656},
  pages={427--430},
  year={2023},
  publisher={American Association for the Advancement of Science}
}

@article{barbarino2015magnetic,
  title={Magnetic crystals and helical liquids in alkaline-earth fermionic gases},
  author={Barbarino, Simone and Taddia, Luca and Rossini, Davide and Mazza, Leonardo and Fazio, Rosario},
  journal={Nature communications},
  volume={6},
  number={1},
  pages={8134},
  year={2015},
  publisher={Nature Publishing Group UK London}
}

@article{yan2015topological,
  title={Topological superfluid and Majorana zero modes in synthetic dimension},
  author={Yan, Zhongbo and Wan, Shaolong and Wang, Zhong},
  journal={Scientific reports},
  volume={5},
  number={1},
  pages={15927},
  year={2015},
  publisher={Nature Publishing Group UK London}
}

@article{cornfeld2015chiral,
  title={Chiral currents in one-dimensional fractional quantum Hall states},
  author={Cornfeld, Eyal and Sela, Eran},
  journal={Physical Review B},
  volume={92},
  number={11},
  pages={115446},
  year={2015},
  publisher={APS}
}

@article{barbarino2016synthetic,
  title={Synthetic gauge fields in synthetic dimensions: interactions and chiral edge modes},
  author={Barbarino, Simone and Taddia, Luca and Rossini, Davide and Mazza, Leonardo and Fazio, Rosario},
  journal={New Journal of Physics},
  volume={18},
  number={3},
  pages={035010},
  year={2016},
  publisher={IOP Publishing}
}

@article{zeng2015charge,
  title={Charge pumping of interacting fermion atoms in the synthetic dimension},
  author={Zeng, Tian-Sheng and Wang, Ce and Zhai, Hui},
  journal={Physical review letters},
  volume={115},
  number={9},
  pages={095302},
  year={2015},
  publisher={APS}
}

@article{taddia2017topological,
  title={Topological fractional pumping with alkaline-earth-like atoms in synthetic lattices},
  author={Taddia, Luca and Cornfeld, Eyal and Rossini, Davide and Mazza, Leonardo and Sela, Eran and Fazio, Rosario},
  journal={Physical review letters},
  volume={118},
  number={23},
  pages={230402},
  year={2017},
  publisher={APS}
}

@article{junemann2017exploring,
  title={Exploring interacting topological insulators with ultracold atoms: The synthetic Creutz-Hubbard model},
  author={J{\"u}nemann, J and Piga, A and Ran, S-J and Lewenstein, M and Rizzi, M and Berm{\'u}dez, Alejandro},
  journal={Physical Review X},
  volume={7},
  number={3},
  pages={031057},
  year={2017},
  publisher={APS}
}

@article{ghosh2015baryon,
  title={Baryon squishing in synthetic dimensions by effective SU (M) gauge fields},
  author={Ghosh, Sudeep Kumar and Yadav, Umesh K and Shenoy, Vijay B},
  journal={Physical Review A},
  volume={92},
  number={5},
  pages={051602},
  year={2015},
  publisher={APS}
}

@article{calvanese2017laughlin,
  title={Laughlin-like states in bosonic and fermionic atomic synthetic ladders},
  author={Calvanese Strinati, Marcello and Cornfeld, Eyal and Rossini, Davide and Barbarino, Simone and Dalmonte, Marcello and Fazio, Rosario and Sela, Eran and Mazza, Leonardo},
  journal={Physical Review X},
  volume={7},
  number={2},
  pages={021033},
  year={2017},
  publisher={APS}
}

@article{saito2017devil,
  title={Devil's staircases in synthetic dimensions and gauge fields},
  author={Saito, Takeshi Y and Furukawa, Shunsuke},
  journal={Physical Review A},
  volume={95},
  number={4},
  pages={043613},
  year={2017},
  publisher={APS}
}

@article{mamaev2022resonant,
  title={Resonant dynamics of strongly interacting SU (n) fermionic atoms in a synthetic flux ladder},
  author={Mamaev, Mikhail and Bilitewski, Thomas and Sundar, Bhuvanesh and Rey, Ana Maria},
  journal={PRX Quantum},
  volume={3},
  number={3},
  pages={030328},
  year={2022},
  publisher={APS}
}

@article{barbiero2023frustrated,
  title={Frustrated magnets without geometrical frustration in bosonic flux ladders},
  author={Barbiero, Luca and Cabedo, Josep and Lewenstein, Maciej and Tarruell, Leticia and Celi, Alessio},
  journal={Physical Review Research},
  volume={5},
  number={4},
  pages={L042008},
  year={2023},
  publisher={APS}
}

@article{buser2020interacting,
  title={Interacting bosonic flux ladders with a synthetic dimension: Ground-state phases and quantum quench dynamics},
  author={Buser, Maximilian and Hubig, Claudius and Schollw{\"o}ck, Ulrich and Tarruell, Leticia and Heidrich-Meisner, Fabian},
  journal={Physical Review A},
  volume={102},
  number={5},
  pages={053314},
  year={2020},
  publisher={APS}
}

@article{lauria2022,
  title = {Experimental Realization of a Fermionic Spin-Momentum Lattice},
  author = {Lauria, Paul and Kuo, Wei-Ting and Cooper, Nigel R. and Barreiro, Julio T.},
  journal = {Phys. Rev. Lett.},
  volume = {128},
  issue = {24},
  pages = {245301},
  numpages = {7},
  year = {2022},
  month = {Jun},
  publisher = {American Physical Society},
  doi = {10.1103/PhysRevLett.128.245301},
  url = {https://link.aps.org/doi/10.1103/PhysRevLett.128.245301}
}

@article{wang2022observation,
  title={Observation of interaction-induced mobility edge in an atomic Aubry-Andr{\'e} wire},
  author={Wang, Yunfei and Zhang, Jia-Hui and Li, Yuqing and Wu, Jizhou and Liu, Wenliang and Mei, Feng and Hu, Ying and Xiao, Liantuan and Ma, Jie and Chin, Cheng and others},
  journal={Physical Review Letters},
  volume={129},
  number={10},
  pages={103401},
  year={2022},
  publisher={APS}
}

@article{li2022aharonov,
  title={Aharonov-Bohm caging and inverse Anderson transition in ultracold atoms},
  author={Li, Hang and Dong, Zhaoli and Longhi, Stefano and Liang, Qian and Xie, Dizhou and Yan, Bo},
  journal={Physical Review Letters},
  volume={129},
  number={22},
  pages={220403},
  year={2022},
  publisher={APS}
}

@article{zeng2024transition,
  title={Transition from Flat-Band Localization to Anderson Localization in a One-Dimensional Tasaki Lattice},
  author={Zeng, Chao and Shi, Yue-Ran and Mao, Yi-Yi and Wu, Fei-Fei and Xie, Yan-Jun and Yuan, Tao and Zhang, Wei and Dai, Han-Ning and Chen, Yu-Ao and Pan, Jian-Wei},
  journal={Physical Review Letters},
  volume={132},
  number={6},
  pages={063401},
  year={2024},
  publisher={APS}
}

@article{fabre2024atomic,
  title={Atomic topological quantum matter using synthetic dimensions},
  author={Fabre, Aur{\'e}lien and Nascimbene, Sylvain},
  journal={Europhysics Letters},
  volume={145},
  number={6},
  pages={65001},
  year={2024},
  publisher={IOP Publishing}
}

@article{wei2014theory,
  title={Theory of bosons in two-leg ladders with large magnetic fields},
  author={Wei, Ran and Mueller, Erich J},
  journal={Physical Review A},
  volume={89},
  number={6},
  pages={063617},
  year={2014},
  publisher={APS}
}

@article{harper1955single,
  title={Single band motion of conduction electrons in a uniform magnetic field},
  author={Harper, Philip George},
  journal={Proceedings of the Physical Society. Section A},
  volume={68},
  number={10},
  pages={874},
  year={1955},
  publisher={IOP Publishing}
}

@article{hofstadter1976energy,
  title={Energy levels and wave functions of Bloch electrons in rational and irrational magnetic fields},
  author={Hofstadter, Douglas R},
  journal={Physical review B},
  volume={14},
  number={6},
  pages={2239},
  year={1976},
  publisher={APS}
}

@article{PhysRevLett.81.3108,
title = {Cold Bosonic Atoms in Optical Lattices},
  author = {Jaksch, D. and Bruder, C. and Cirac, J. I. and Gardiner, C. W. and Zoller, P.},
  journal = {Phys. Rev. Lett.},
  volume = {81},
  issue = {15},
  pages = {3108--3111},
  numpages = {0},
  year = {1998},
  month = {Oct},
  publisher = {American Physical Society},
  doi = {10.1103/PhysRevLett.81.3108},
  url = {https://link.aps.org/doi/10.1103/PhysRevLett.81.3108}
}

@article{Atala_2014,
   title={Observation of chiral currents with ultracold atoms in bosonic ladders},
   volume={10},
   ISSN={1745-2481},
   url={http://dx.doi.org/10.1038/nphys2998},
   DOI={10.1038/nphys2998},
   number={8},
   journal={Nature Physics},
   publisher={Springer Science and Business Media LLC},
   author={Atala, Marcos and Aidelsburger, Monika and Lohse, Michael and Barreiro, Julio T. and Paredes, Belén and Bloch, Immanuel},
   year={2014},
   month=jul, pages={588–593} }

@misc{rudner2020floquetengineershandbook,
      title={The Floquet Engineer's Handbook}, 
      author={Mark S. Rudner and Netanel H. Lindner},
      year={2020},
      eprint={2003.08252},
      archivePrefix={arXiv},
      primaryClass={cond-mat.mes-hall},
      url={https://arxiv.org/abs/2003.08252}, 
}

@article{Gauthier_2016,
   title={Direct imaging of a digital-micromirror device for configurable microscopic optical potentials},
   volume={3},
   ISSN={2334-2536},
   url={http://dx.doi.org/10.1364/OPTICA.3.001136},
   DOI={10.1364/optica.3.001136},
   number={10},
   journal={Optica},
   publisher={Optica Publishing Group},
   author={Gauthier, G. and Lenton, I. and McKay Parry, N. and Baker, M. and Davis, M. J. and Rubinsztein-Dunlop, H. and Neely, T. W.},
   year={2016},
   month=oct, pages={1136} }

@article{Amico_2021,
   title={Roadmap on Atomtronics: State of the art and perspective},
   volume={3},
   ISSN={2639-0213},
   url={http://dx.doi.org/10.1116/5.0026178},
   DOI={10.1116/5.0026178},
   number={3},
   journal={AVS Quantum Science},
   publisher={American Vacuum Society},
   author={Amico, L. and others},
   year={2021},
   month=aug }

@article{PhysRevA.77.033629,
  title = {Vortex lattices of bosons in deep rotating optical lattices},
  author = {Goldbaum, Daniel S. and Mueller, Erich J.},
  journal = {Phys. Rev. A},
  volume = {77},
  issue = {3},
  pages = {033629},
  numpages = {6},
  year = {2008},
  month = {Mar},
  publisher = {American Physical Society},
  doi = {10.1103/PhysRevA.77.033629},
  url = {https://link.aps.org/doi/10.1103/PhysRevA.77.033629}
}

@article{piraud2015vortex,
  title={Vortex and Meissner phases of strongly interacting bosons on a two-leg ladder},
  author={Piraud, Marie and Heidrich-Meisner, Fabian and McCulloch, Ian P and Greschner, Sebastian and Vekua, Temo and Schollwoeck, Ulrich},
  journal={Physical Review B},
  volume={91},
  number={14},
  pages={140406},
  year={2015},
  publisher={APS}
}

@article{uchino2016analytical,
  title={Analytical approach to a bosonic ladder subject to a magnetic field},
  author={Uchino, Shun},
  journal={Physical Review A},
  volume={93},
  number={5},
  pages={053629},
  year={2016},
  publisher={APS}
}

@article{petrescu2015chiral,
  title={Chiral Mott insulators, Meissner effect, and Laughlin states in quantum ladders},
  author={Petrescu, Alexandru and Le Hur, Karyn},
  journal={Physical Review B},
  volume={91},
  number={5},
  pages={054520},
  year={2015},
  publisher={APS}
}

@article{greschner2016symmetry,
  title={Symmetry-broken states in a system of interacting bosons on a two-leg ladder with a uniform Abelian gauge field},
  author={Greschner, Sebastian and Piraud, M and Heidrich-Meisner, F and McCulloch, IP and Schollw{\"o}ck, Ulrich and Vekua, T},
  journal={Physical Review A},
  volume={94},
  number={6},
  pages={063628},
  year={2016},
  publisher={APS}
}

@article{kelecs2015mott,
  title={Mott transition in a two-leg Bose-Hubbard ladder under an artificial magnetic field},
  author={Kele{\c{s}}, Ahmet and Oktel, M{\"O}},
  journal={Physical Review A},
  volume={91},
  number={1},
  pages={013629},
  year={2015},
  publisher={APS}
}

@article{qiao2021quantum,
  title={Quantum phases of interacting bosons on biased two-leg ladders with magnetic flux},
  author={Qiao, Xin and Zhang, Xiao-Bo and Jian, Yue and Zhang, Ai-Xia and Yu, Zi-Fa and Xue, Ju-Kui},
  journal={Physical Review A},
  volume={104},
  number={5},
  pages={053323},
  year={2021},
  publisher={APS}
}

@article{uchino2015population,
  title={Population-imbalance instability in a Bose-Hubbard ladder in the presence of a magnetic flux},
  author={Uchino, Shun and Tokuno, Akiyuki},
  journal={Physical Review A},
  volume={92},
  number={1},
  pages={013625},
  year={2015},
  publisher={APS}
}

@article{natu2015bosons,
  title={Bosons with long-range interactions on two-leg ladders in artificial magnetic fields},
  author={Natu, Stefan S},
  journal={Physical Review A},
  volume={92},
  number={5},
  pages={053623},
  year={2015},
  publisher={APS}
}

@article{kolovsky2017bogoliubov,
  title={Bogoliubov depletion of the fragmented condensate in the bosonic flux ladder},
  author={Kolovsky, Andrey R},
  journal={Physical Review A},
  volume={95},
  number={3},
  pages={033622},
  year={2017},
  publisher={APS}
}

@article{alexander1983superconductivity,
  title={Superconductivity of networks. A percolation approach to the effects of disorder},
  author={Alexander, S},
  journal={Physical Review B},
  volume={27},
  number={3},
  pages={1541},
  year={1983},
  publisher={APS}
}

@article{aidelsburger,
  title = {Realization of the Hofstadter Hamiltonian with Ultracold Atoms in Optical Lattices},
  author = {Aidelsburger, M. and Atala, M. and Lohse, M. and Barreiro, J. T. and Paredes, B. and Bloch, I.},
  journal = {Phys. Rev. Lett.},
  volume = {111},
  issue = {18},
  pages = {185301},
  numpages = {5},
  year = {2013},
  month = {Oct},
  publisher = {American Physical Society},
  doi = {10.1103/PhysRevLett.111.185301},
  url = {https://link.aps.org/doi/10.1103/PhysRevLett.111.185301}
}

@article{impertro2025strongly,
  title={Strongly interacting Meissner phases in large bosonic flux ladders},
  author={Impertro, Alexander and Huh, SeungJung and Karch, Simon and Wienand, Julian F and Bloch, Immanuel and Aidelsburger, Monika},
  journal={Nature Physics},
  pages={1--7},
  year={2025},
  publisher={Nature Publishing Group UK London}
}

\end{document}